%% file: main.tex
\documentclass[twocolumn,amsmath,amssymb,a4paper,prb,superscriptaddress,floatfix]{revtex4-2}
\usepackage{graphicx}
\usepackage{tikz}
\usetikzlibrary{arrows.meta, positioning, calc}
\usepackage{bm}
\usepackage{url}
\usepackage{xcolor}
\usepackage[hidelinks]{hyperref}

\newif\ifnotes
\notestrue
\newcommand{\note}[1]{\ifnotes\par\textcolor{gray}{\textit{[#1]}}\par\fi}

\newcommand{\Fvib}{F_\mathrm{vib}}
\newcommand{\Fel}{F_\mathrm{el}}

\newcommand{\kB}{k_\mathrm{B}}

\input{tables/alpha-t0}
\input{tables/einstein-zero}

\begin{document}

\title{Axial thermal expansion from free-energy minimization with
  temperature-dependent force constants in phonopy: a technical report, with
  $\alpha$- and $\omega$-Ti as the worked example}

\author{Atsushi Togo}
\email{togo.atsushi@nims.go.jp}
\affiliation{Center for Basic Research on Materials, National Institute for Materials Science, Tsukuba, Ibaraki 305-0047, Japan}

\begin{abstract}
  In a hexagonal crystal the $a$ and $c$ axes change their lengths at
  different rates with increasing temperature. The thermal expansion of
  such a crystal is therefore described by two coefficients, one for $a$
  and one for $c$. The equilibrium lattice parameters at each temperature
  are found by varying the lattice parameters to minimize the Helmholtz
  free energy, and the coefficients are their logarithmic derivatives with
  respect to temperature. This report describes a procedure for this
  minimization in the form implemented in phonopy. In this procedure, the
  free energy is computed with harmonic force constants at a finite set of
  values of $a$ and $c$, and a free-energy surface over $a$ and $c$ is
  fitted to these values. The report also describes a variant in which the
  harmonic force constants are replaced by temperature-dependent force
  constants, obtained from a self-consistent harmonic approximation with
  the forces computed by a machine-learning potential fitted at each of
  these values of $a$ and $c$. The $\alpha$ and $\omega$ phases of
  titanium are the worked example. The $c$ axis of $\alpha$-Ti shows
  negative thermal expansion at low temperature, and the $c$ axis of
  $\omega$-Ti does not. The calculations with harmonic force constants and
  with temperature-dependent force constants both give the negative
  thermal expansion of the $c$ axis in $\alpha$-Ti and its absence in
  $\omega$-Ti. The difference between the two calculations is most
  noticeable for the $c$ axis of $\alpha$-Ti. Computing the axial thermal
  expansion coefficients is harder in principle than computing the
  volumetric thermal expansion coefficient. The two axes are coupled in the
  free energy. For example, the free energy increases when $a$ is
  lengthened, and part of this increase is cancelled when $c$ is shortened
  at the same time. The thermal effect that lengthens $a$ can also act to
  shorten $c$, as it does in both phases of titanium. In that case two
  terms of opposite sign make up the thermal expansion coefficient of the
  $c$ axis. One is the shortening of $c$ caused by the thermal effect on
  $a$. The other is the lengthening of $c$ caused by the thermal effect on
  $c$ itself. The two terms partly cancel, so the coefficient can be much
  smaller than either term. A small error in either term therefore gives a
  large relative error in the coefficient. In the volumetric thermal
  expansion coefficient the corresponding terms have the same sign and do
  not cancel. The report measures how much each setting of the procedure
  changes the thermal expansion coefficients, with the $c$ axis of
  $\alpha$-Ti as the most sensitive case.
\end{abstract}

\maketitle

\section{Introduction}

In phonopy, thermal expansion is computed in the quasi-harmonic
approximation. The Helmholtz free energy, called simply the free energy in
this report, is evaluated from first-principles phonon calculations at
several volumes. At each temperature, an equation of state fitted to these
free energies is minimized over the volume, which gives the equilibrium
volume.\cite{phonopy-QHA-max,phonopy,phonopy-phono3py} The volumetric
thermal expansion coefficient is computed from the equilibrium volumes at
the different temperatures. In a hexagonal crystal the symmetry requires the
basis vectors $\mathbf{a}$ and $\mathbf{b}$ to have the same length,
$|\mathbf{a}| = |\mathbf{b}|$, and fixes the angles between the basis
vectors. Only the lattice parameters $a = |\mathbf{a}|$ and $c =
|\mathbf{c}|$ can therefore change independently. Lattice parameters that
can change independently are called free lattice parameters in this report.
The $a$ and $c$ axes change their lengths at different rates with increasing
temperature, and the thermal expansion is described by one coefficient for
each axis. In an orthorhombic crystal $a$, $b$ and $c$ are all free, and
three coefficients are needed.

The minimization over the volume does not determine the axial thermal
expansion coefficients from the free energy. The shape of the unit cell at
each volume is fixed before the minimization, when the unit cells are made.
Typical choices are to relax the unit cell at each volume with the static
energy, or to keep the axial ratio $c/a$ fixed. Whichever choice is made,
$c/a$ at each volume is chosen when the unit cells are made. When the free
energy is minimized over $a$ and $c$ instead, the shape of the unit cell at
each temperature is also determined by the free energy.

Minimizing over the free lattice parameters makes the calculation larger.
The free energy has to be computed at many combinations of $a$ and $c$,
rather than at several volumes along one path, and it has to be
differentiated in more than one direction. For each combination of $a$ and
$c$, a reference cell is strained to those lattice parameters. The resulting unit cell is called a strained cell in this report.

In the zero static internal stress approximation,\cite{allan-zsisa} the
internal coordinates of the atoms are relaxed with the static energy at each
set of lattice parameters, and the free energy is minimized over the lattice
parameters only. This approximation has also been applied to
thermoelasticity,\cite{gong-beryllium,gong-osmium} and it has been extended
to anisotropic thermal expansion.\cite{rostami-anisotropic} These works aim
to reduce the computational cost. This report does not aim at a low cost,
and it keeps the procedure simple instead. The free energy is computed at
every strained cell, and a surface is fitted to these values. As in the zero
static internal stress approximation, the internal coordinates of the atoms
are relaxed with the static energy only. Relaxing them in the free energy at
finite temperature is not implemented in phonopy at the time of writing. The
$\alpha$ and $\omega$ phases of titanium have no free internal coordinates,
so this does not affect the results of this report. This report describes
how the free energy is minimized over the lattice parameters with phonopy,
and what is needed to do this with temperature-dependent force constants.

In the quasi-harmonic approximation, the vibrational free energy of each
strained cell is computed with force constants that do not change with
temperature. For a crystal with strong anharmonicity this approximation is
not sufficient, and temperature-dependent force constants have to be used.
Several methods give such force constants, for example the self-consistent
phonon
method.\cite{Tadano-2015,PhysRevB.105.064112,PhysRevB.107.134119,PhysRevB.106.224104}
This report uses the stochastic
self-consistent harmonic approximation
(SSCHA).\cite{Errea-SSCHA-2013,Errea-SSCHA-2014,Bianco-SSCHA-2017,Monacelli-SSCHA-2021,van-Roekeghem-2020}
The SSCHA needs the forces of thousands of supercells with displacements
at each strained cell and each temperature. Computing those forces from first
principles at every strained cell is too expensive. The procedure proposed
in phonopy therefore fits a machine-learning potential at each strained cell
to the first-principles forces of a smaller set of supercells, and the SSCHA
computes the forces with that potential. With these potentials, the SSCHA
can be applied at every strained cell.

This report describes both calculations, with harmonic force constants and
with temperature-dependent force constants from the SSCHA, as they are
implemented in phonopy version~4.6. Both calculations are applied to the
$\alpha$ and $\omega$ phases of titanium. Both phases are hexagonal, so
both have two free lattice parameters. Both phases consist of the same
element and are computed with approximately the same settings, so the
results for the two phases differ mainly because of the crystal
structure. The axial thermal
expansions of the two phases differ qualitatively. The $c$ axis of
$\alpha$-Ti shows negative thermal expansion below about 110~K, and the $c$
axis of $\omega$-Ti expands at every temperature. The calculations with
harmonic force constants and with temperature-dependent force constants
both give this difference. The computed axial thermal expansion
coefficients are not compared with experiment in this report. That comparison is left to other work.

The axial thermal expansion coefficients are obtained from the free energies
of the chosen strained cells in three steps. In the first step, a
free-energy surface over $a$ and $c$ is fitted to the free energies of the
strained cells at each temperature. In the second step, the equilibrium
lattice parameters $a$ and $c$ at each temperature are given by the minimum
of this surface. In the third step, a smooth function of temperature is
fitted to the equilibrium values of $a$, and another to those of $c$. The
axial thermal expansion coefficients are the logarithmic derivatives of
these functions with respect to temperature. Any smooth function of
temperature can be used. In this report each function is a sum of Einstein
terms (Sec.~\ref{sec:extraction}), and this fit is called the Einstein fit.

Sec.~\ref{sec:choices} lists the settings of this procedure that the user
chooses, and measures how much each of them changes the axial thermal
expansion coefficients.

\section{Free energy and its minimization}

\subsection{The free energy as a function of the lattice parameters}

In the implementation of phonopy, the Helmholtz free energy is written as a
function of the lattice parameters $\bm{x}$ and the temperature $T$,
\begin{equation}
  F(\bm{x}; T) = U(\bm{x}) + \Fel(\bm{x}; T) + \Fvib(\bm{x}; T),
  \label{eq:total}
\end{equation}
where $U$ is the static internal energy of the electronic ground state,
$\Fel$ is the electronic free energy of the thermally occupied electronic
states, and $\Fvib$ is the vibrational free energy. The three terms are
given per primitive cell. For a hexagonal crystal $\bm{x} = (a, c)$. Both
phases of titanium in this report are hexagonal.

At each temperature the free energy is minimized with respect to the lattice
parameters,
\begin{equation}
  \bm{x}(T) = \arg\min_{\bm{x}} F(\bm{x}; T),
  \label{eq:minimize}
\end{equation}
where $\bm{x}(T)$ are the equilibrium lattice parameters, with components $a(T)$ and $c(T)$. Without the argument $T$, $\bm{x}$,
$a$ and $c$ denote the lattice parameters as variables. The axial thermal expansion
coefficients are the logarithmic derivatives of the equilibrium lattice
parameters,
\begin{equation}
  \begin{split}
    \alpha_a(T) &= \frac{1}{a(T)}\frac{da(T)}{dT} = \frac{d\ln a(T)}{dT}, \\
    \alpha_c(T) &= \frac{1}{c(T)}\frac{dc(T)}{dT} = \frac{d\ln c(T)}{dT}.
  \end{split}
  \label{eq:alpha}
\end{equation}

In the volume quasi-harmonic approximation, Eq.~(\ref{eq:minimize}) is
applied with $\bm{x} = V$, and $U(V) + \Fvib(V; T) + pV$ is minimized over
the volume.\cite{phonopy-QHA-max} The unit cells at the different volumes are made before the minimization, and their $c/a$ is fixed when they are made.
The minimization then determines only the volume at each temperature. When
the free energy is minimized over $a$ and $c$ instead, both $a(T)$ and
$c(T)$ are determined by the free energy.

\subsection{The surface and its minimum}
\label{sec:minimization}

The minimization of Eq.~(\ref{eq:minimize}) needs the free energy
$F(\bm{x}; T)$ at every $\bm{x}$, as a surface over the lattice parameters.
This surface is not computed directly. The free energy is computed at a
finite number of chosen values $\bm{x}_j$, $j = 1 \dots N_x$, and a surface
is fitted to those values. The unit cell at each $\bm{x}_j$ is a strained cell,
made by straining a reference cell to the lattice parameters $\bm{x}_j$.
Section~\ref{sec:free-parameters} describes how the strained cells are made
under the crystal symmetry.

The three terms of Eq.~(\ref{eq:total}) are computed at every strained cell
$\bm{x}_j$, and their sum is fitted by least squares with a polynomial of
total degree $n$ in the free lattice parameters,
\begin{equation}
  P(\bm{x}) = \sum_{|\bm{p}| \le n} c_{\bm{p}}\, u_1^{p_1} u_2^{p_2},
  \qquad
  \bm{u} = \frac{\bm{x} - \bar{\bm{x}}}{\Delta \bm{x}},
  \label{eq:poly}
\end{equation}
where $\bm{p} = (p_1, p_2)$ are non-negative integers with
$|\bm{p}| = p_1 + p_2$, $\bar{\bm{x}}$ is the mean of the $\bm{x}_j$, and
$\Delta\bm{x}$ is half the range that the $\bm{x}_j$ cover along each
lattice parameter. The region that the $\bm{x}_j$ cover is called the fitting region in this
report. Inside it the fitted surface interpolates the computed free
energies, and outside it the surface is an extrapolation. The scaling by $\Delta\bm{x}$ does not change the fitted surface, and it
keeps the least-squares fit well conditioned (Sec.~\ref{sec:surface}). For $d$
free lattice parameters the polynomial has $\binom{n+d}{n}$ coefficients,
which is ten for $d = 2$ and $n = 3$. The number of strained cells $N_x$
must be at least the number of coefficients.

The least-squares fit determines the $c_{\bm{p}}$ of Eq.~(\ref{eq:poly})
from the design matrix $X$ of the fit, whose entry for the strained cell
$\bm{x}_j$ and the exponent pair $\bm{p}$ is
\begin{equation}
  X_{j\bm{p}} = u_{1j}^{p_1} u_{2j}^{p_2},
  \label{eq:design}
\end{equation}
where $\bm{u}_j$ is computed from $\bm{x}_j$ as in Eq.~(\ref{eq:poly}).
When $U$, $\Fel$ and $\Fvib$ are computed at the same strained cells and
fitted at the same degree, their three fits share one design matrix. The
least-squares coefficients are then a linear function of the fitted
values, so fitting $U$, $\Fel$ and $\Fvib$ of Eq.~(\ref{eq:total})
separately and adding the three polynomials gives the same surface as
fitting their sum $F$. Fitting the terms separately would allow $U$ and $\Fel$ to be
computed at more strained cells than $\Fvib$. This would be useful because
the static calculation of a strained cell uses the unit cell, and it costs
about three orders of magnitude less than the supercell calculations of
$\Fvib$ at the same strained cell. However, the implementation used here
computes all three terms at the same strained cells.

The fitted polynomial is minimized at each temperature by the
Broyden--Fletcher--Goldfarb--Shanno (BFGS) method,\cite{nocedal-wright} as
implemented in \texttt{scipy.optimize.minimize} of SciPy.\cite{scipy} The
iteration starts from the mean $\bar{\bm{x}}$ of the $\bm{x}_j$. Phonopy
computes the gradient of the polynomial analytically from its coefficients
and passes it to the minimizer.
The iteration is stopped when the largest component of the gradient is
smaller than $10^{-9}$~eV/\AA. This threshold is four orders of magnitude
tighter than the default of SciPy's BFGS at the time of writing,
$10^{-5}$ in SciPy version~1.18. The surface is a polynomial with
an analytic gradient, so the tight threshold adds little computing time.

At the minimum of the polynomial the gradient is zero. At the point where
the iteration stops, the gradient is not exactly zero, and its largest
component is below the threshold. The remaining gradient divided by the
curvature of the surface at that point is the distance to the minimum. This
distance is called the position error in this report, and it is at most
about the gradient threshold divided by the curvature. The curvature along
$c$ is about one seventh of the curvature along $a$ ($1.8$ against
$12.5$~eV/\AA$^2$ for $\alpha$-Ti at 300~K), so the position error is
largest in $c$. The minimization is carried out separately at each
temperature, so the position error differs from one temperature to the next.
The position error therefore appears as scatter in the equilibrium lattice parameters $\bm{x}(T)$, and the Einstein fit is made to $\bm{x}(T)$
including this scatter.

With a threshold this tight, the line search of the BFGS iteration
sometimes stops before the gradient is smaller than $10^{-9}$~eV/\AA,
because floating-point precision is lost. SciPy then reports a failure.
Convergence is therefore judged from the gradient at the point where the
iteration stopped, and not only from what SciPy reports. If the largest
component of that gradient is smaller than $10^{-6}$~eV/\AA{}, the
minimization is accepted as converged.

The position error has to be smaller than the differences in the lattice
parameters that this report measures. At $10^{-9}$~eV/\AA{} the position
error of the minimization in $c$ of $\alpha$-Ti is $6 \times 10^{-7}$~m\AA. This is a
sixteen-hundredth of the scatter of $c(T)$ about its Einstein
fit, $9 \times 10^{-4}$~m\AA{} (Sec.~\ref{sec:terms}). At SciPy's
default of $10^{-5}$~eV/\AA{} the position error would be
$6 \times 10^{-3}$~m\AA, six times that scatter. The computed $\bm{x}(T)$ would then show where the iteration stopped rather than where
the free energy has its minimum. The comparison of the Einstein fits in
Sec.~\ref{sec:terms} uses residuals as small as $9 \times 10^{-4}$~m\AA, which is smaller than the
position error at the default threshold.

The curvature used in these estimates comes almost entirely from $U$. The
static energy alone gives $12.76$ and $1.85$~eV~\AA$^{-2}$ along $a$ and
$c$, against $12.48$ and $1.77$ for the total free energy, as
Sec.~\ref{sec:results} shows. The two agree to within a few per cent
because $\Fvib$ and $\Fel$ have small curvatures over a fitting region this narrow.
The slopes of $\Fvib$ are not small, and they move the minimum away from
the minimum of $U$, as Sec.~\ref{sec:results} also shows.

\subsection{The vibrational free energy}
\label{sec:fvib}

In this report $\Fvib$ is computed in two ways. One way uses the harmonic
approximation, and the other uses the SSCHA. In both ways $\Fvib$ is
obtained per primitive cell at each strained cell and each temperature, and
these values enter Eq.~(\ref{eq:total}). The fit of Eq.~(\ref{eq:poly})
uses only these values, so the rest of the analysis is the same for both
ways. In Fig.~\ref{fig:workflow} the harmonic $\Fvib$ is computed from the
dataset of the left column, and the SSCHA $\Fvib$ comes from the right
column. Both are analysed by the same command,
\texttt{phonopy-anisotropic-qha}.

The first is the harmonic free energy of the force constants $\Phi$ of a
strained cell,
\begin{equation}
  \begin{split}
    F_\mathrm{harm}(\Phi; T) = \frac{1}{N_{\bm{q}}}\sum_{\bm{q}\nu}
    \Bigl[ &\frac{\hbar\omega_{\bm{q}\nu}}{2} \\
    &+ \kB T \ln\!\left(1 - e^{-\hbar\omega_{\bm{q}\nu}/\kB T}\right) \Bigr],
  \end{split}
  \label{eq:fharm}
\end{equation}
where $\omega_{\bm{q}\nu}$ is the frequency of the phonon mode $\nu$ at the wave
vector $\bm{q}$, obtained from the dynamical matrix of $\Phi$, and
$N_{\bm{q}}$ is the number of sampled $\bm{q}$ points. In the
quasi-harmonic approximation, $\Phi$ is obtained from the forces on the
atoms of supercells with displacements. These forces are often computed by a
density-functional-theory code. The program that computes the forces is called
the calculator in this report. Here the calculator is VASP.

The second way uses the free energy of the SSCHA, in the form implemented
in phonopy.
In the SSCHA the force constants $\Phi$ are the parameters
of a trial harmonic system. They define a harmonic density matrix $\tilde{\rho}_\Phi$ at $T$. The SSCHA free energy is
\begin{equation}
  \mathcal{F}(\Phi; T) = F_\mathrm{harm}(\Phi; T)
  + \frac{1}{n_\mathrm{cell}} \left(
    \langle V \rangle_{\tilde{\rho}_\Phi}
    - \langle \tilde{V}_\Phi \rangle_{\tilde{\rho}_\Phi} \right),
  \label{eq:fsscha}
\end{equation}
where $\mathcal{F}(\Phi; T)$ is the SSCHA free energy of one strained cell
for the trial force constants $\Phi$, per primitive cell as
$F_\mathrm{harm}$ is. The averages are taken over supercells with displacements:
$V$ is the potential energy of the supercell, measured from its value at
the undisplaced positions, $\tilde{V}_\Phi$ is the harmonic potential
energy of the supercell for $\Phi$, and $n_\mathrm{cell}$ is the number of
primitive cells in the supercell. The SSCHA looks for the force
constants $\Phi_\mathrm{sc}$ at which $\mathcal{F}(\Phi; T)$ is stationary
with respect to $\Phi$, called the self-consistent force constants, and
$\mathcal{F}(\Phi_\mathrm{sc}; T)$ is $\Fvib$ of the strained cell.

In practice the two averages are estimated together from $N$ supercells
whose atoms are displaced at random according to $\tilde{\rho}_\Phi$,
\begin{equation}
  \langle V \rangle_{\tilde{\rho}_\Phi}
  - \langle \tilde{V}_\Phi \rangle_{\tilde{\rho}_\Phi}
  \approx \frac{1}{N}\sum_{i=1}^{N}
  \left[ V(\bm{u}_i) - \tfrac{1}{2}\bm{u}_i^{\mathsf T}\Phi\bm{u}_i \right],
  \label{eq:sscha-estimate}
\end{equation}
where $\bm{u}_i$ is the displacement of the atoms of the $i$th supercell.
The average of $\tilde{V}_\Phi$ could also be computed exactly from the
frequencies and eigenvectors of $\Phi$.\cite{Errea-SSCHA-2014} Taking it over the same supercells
as $V$ makes the difference of the two averages converge more stably with
the number of supercells.\cite{Togo-IXS-KCl-2022}

An SSCHA run computes $\Fvib$ at one strained cell and one temperature. The
run is a sequence of iterations. The runs at every strained cell and every
temperature together are called a sweep. Iteration $k$ of a run draws $N$
supercells with displacements from $\tilde\rho_{\Phi_k}$. The displacements
are drawn in the same way as the training supercells of
Sec.~\ref{sec:training}, with $\Phi_k$ in place of the harmonic force
constants. The energies and forces of these supercells are computed with the machine-learning potential of the
strained cell, which is described in Sec.~\ref{sec:tdfc}. The free energy of
iteration $k$ is
\begin{equation}
  F^{(k)}(T) = F_\mathrm{harm}(\Phi_k; T) + \frac{1}{N}\sum_{i=1}^{N} \Delta^{(k)}_i,
  \label{eq:fiteration}
\end{equation}
with
\begin{equation}
  \Delta^{(k)}_i = \frac{1}{n_\mathrm{cell}}
        \left( E_i - E_0
        - \tfrac{1}{2}\bm{u}^{(k)\mathsf T}_i\Phi_k\bm{u}^{(k)}_i \right),
  \label{eq:anharmonic}
\end{equation}
where $E_i$ is the energy of the $i$th supercell and $E_0$ that of the
supercell without displacements. The quantity $\Delta^{(k)}_i$, the anharmonic
energy of the $i$th supercell per primitive cell, is the bracket of
Eq.~(\ref{eq:sscha-estimate}) divided by $n_\mathrm{cell}$, with
$\bm{u}_i = \bm{u}^{(k)}_i$, $V(\bm{u}_i) = E_i - E_0$ and
$\Phi = \Phi_k$. The force constants $\Phi_{k+1}$ of the next iteration are
then fitted by least squares\cite{symfc} to the displacements
$\bm{u}^{(k)}_i$ and the forces $\bm{f}^{(k)}_i$ of the same $N$
supercells, so that $\bm{f}^{(k)}_i \approx -\Phi_{k+1}\bm{u}^{(k)}_i$. The
free energy $F^{(k)}$ is the SSCHA free energy
of $\Phi_k$, from which the supercells of the iteration were drawn. It is
not the free energy of $\Phi_{k+1}$, which is fitted to those supercells.
If $F^{(k)}$ were assigned to $\Phi_{k+1}$, the harmonic term and the
average of the $\Delta^{(k)}_i$ would belong to different force constants.

The free energy $F^{(k)}$ of Eq.~(\ref{eq:fiteration}) has two terms. The
harmonic term $F_\mathrm{harm}(\Phi_k; T)$ is determined by $\Phi_k$ and
contains no random sampling. The second term is the average of the
$\Delta^{(k)}_i$ over the $N$ supercells of iteration $k$, whose
displacements are drawn at random from $\tilde\rho_{\Phi_k}$. For a given
$\Phi_k$ the statistical error of $F^{(k)}$ therefore comes from this
average alone, and it is
\begin{equation}
  \hat\sigma^{(k)} = \frac{\mathrm{std}(\Delta^{(k)})}{\sqrt{N}},
  \label{eq:sscha-error}
\end{equation}
where $\mathrm{std}(\Delta^{(k)})$ is the standard deviation of the
$\Delta^{(k)}_i$ of the iteration. Equation~(\ref{eq:sscha-error}) treats $\Phi_k$ as exact. It does
not include the error of $\Phi_k$ itself. The force constants $\Phi_k$ were
fitted to the supercells of the previous iteration, so they have the
sampling error of those supercells. An error $\delta\Phi$ in $\Phi_k$ changes the
harmonic term $F_\mathrm{harm}(\Phi_k; T)$. It also changes the average of the $\Delta^{(k)}_i$, because the supercells of that
average are drawn from $\tilde\rho_{\Phi_k}$. Each part changes at first
order in $\delta\Phi$. The SSCHA free energy of
Eq.~(\ref{eq:fsscha}) is stationary with respect to $\Phi$ at
$\Phi_\mathrm{sc}$.\cite{Errea-SSCHA-2014,Bianco-SSCHA-2017}
Therefore the first-order change of the harmonic term and the first-order
change of the average cancel, and $F^{(k)}$ changes only at second order in
$\delta\Phi$. Equation~(\ref{eq:sscha-error}) does not include this
second-order change.

In the first iterations of a run, $\Phi_k$ still changes systematically,
from its starting value toward $\Phi_\mathrm{sc}$. These iterations are
called the transient. After the transient, $\Phi_k$ is close to
$\Phi_\mathrm{sc}$, but it does not converge to $\Phi_\mathrm{sc}$ exactly.
Each iteration fits new force constants to a new sample
of supercells, so $\Phi_k$ fluctuates around $\Phi_\mathrm{sc}$ from one
iteration to the next.

Successive iterations are not strictly independent. The force constants
$\Phi_{k+1}$ are fitted to the supercells of iteration $k$, so $F^{(k+1)}$
depends on those supercells. Because $\mathcal{F}(\Phi; T)$ is stationary with
respect to $\Phi$ at $\Phi_\mathrm{sc}$, as used above for the error of
$\Phi_k$, this dependence is only of second order. The iterations after the transient are therefore
treated as independent samples of the same quantity, and the vibrational
free energy of a strained cell is their mean,
\begin{equation}
  \Fvib(\bm{x}_j; T) = \frac{1}{K}\sum_{k'=1}^{K} F^{(k_0+k')}(T),
  \label{eq:fvib-mean}
\end{equation}
where $k_0$ is the number of iterations in the transient and $K$ is the number
of iterations kept. The iterations of the transient are not included in this
mean.

When the $\hat\sigma^{(k)}$ of the kept iterations are about equal, their
common value is written $\hat\sigma$, and the error of the mean of
Eq.~(\ref{eq:fvib-mean}) is
$\hat\sigma/\sqrt{K} = \mathrm{std}(\Delta)/\sqrt{KN}$, where
$\mathrm{std}(\Delta)$ is the common value of the $\mathrm{std}(\Delta^{(k)})$. The error depends only on the product $KN$. The error therefore does not decide
how a fixed number of supercell calculations should be divided between $K$
and $N$. That division is decided by the other effects of $K$ and $N$. $K$
has to be large enough that iterations remain after the transient. $N$ determines the noise in the force constants of each iteration, and
every quantity other than the free energy is computed from those force
constants. Most of the supercell calculations are therefore spent on a
large $N$, and $K$ is made only large enough that iterations remain after
the transient.

\subsection{The electronic free energy}

The electronic free energy $\Fel$ of a strained cell is obtained from its
electronic density of states $g(E)$. In a density-functional-theory
calculation, $g(E)$ is obtained from the Kohn--Sham eigenvalues computed on
a $k$-point mesh. At each temperature the chemical
potential $\mu$ is determined by the condition that the number of
electrons equals $n_\mathrm{el}$. The energy and the entropy of the
electrons are then computed at $\mu$,
\begin{align}
  N_\mathrm{el}(\mu, T) &= \int g(E) f(E; \mu, T)\, dE = n_\mathrm{el},
  \label{eq:count} \\
  E_\mathrm{el}(T) &= \int g(E)\, E\, f(E; \mu, T)\, dE, \\
  S(T) &= -\kB \int g(E)
            \left[ f\ln f + (1-f)\ln(1-f) \right] dE, \\
  \Fel(T) &= E_\mathrm{el}(T) - T S(T) - E_\mathrm{el}(0),
  \label{eq:fel}
\end{align}
where $f$ is the Fermi--Dirac distribution. The static energy $U$ already
contains the energy of the electrons in their ground state.
$E_\mathrm{el}(T)$ is a band energy, the eigenvalues weighted by their
occupations, and only its change with temperature is used. The last term in
Eq.~(\ref{eq:fel}) measures $\Fel$ from its value at $T = 0$, so that
$\Fel$ contains only the thermal excitation of the electrons and can be
added to $U$ in Eq.~(\ref{eq:total}).

The density of states is integrated by the linear tetrahedron
method\cite{MacDonald-tetrahedron-1979,Blochl-tetrahedron-1994,phonopy-implementation}
on the $k$-point mesh on which the
electronic states were computed. Alternatively, the Fermi--Dirac
occupations of the Kohn--Sham eigenvalues can be summed directly over the
irreducible $k$ points, each with its weight. This sum treats each
eigenvalue as a discrete level. At low temperature only the levels within a
few $\kB T$ of the Fermi level change their occupation, and on a practical
$k$-point mesh this narrow energy range contains few eigenvalues. The sum
then depends on where these few eigenvalues happen to lie, and it converges
slowly with the number of $k$ points. The tetrahedron method interpolates
the eigenvalues linearly between the $k$ points. This gives a continuous
density of states, and the result does not depend on where the individual
eigenvalues lie. Section~\ref{sec:fel-worth}
measures how much the choice between the two methods changes the axial
thermal expansion coefficients.

The integrals are taken over an energy window around the Fermi level, not
over the whole band. Outside the window the occupations are 0 or 1 at every
temperature of the calculation, so the states outside the window do not
change with temperature. Integrating over the whole band would give
$E_\mathrm{el}(T)$ and $E_\mathrm{el}(0)$ as two large numbers whose
difference, $\Fel$, is smaller by many orders of magnitude, and that
difference would be lost in the rounding error. The window keeps these large
constant contributions out of the integrals. The window extends on each side
of the Fermi level by twelve $\kB T$ of the highest temperature, and by at
least 0.5~eV. Inside the window the density of states is evaluated on an
energy grid with a spacing of 0.5~meV. The implementation reads the
eigenvalues from the calculator and builds the density of states itself, so
that this window and this grid can be chosen.

At $T > 0$ the electron count $N_\mathrm{el}(\mu, T)$ of Eq.~(\ref{eq:count}) is a
smooth function of $\mu$. The integral of $N_\mathrm{el}(\mu, T)$ is evaluated on the
energy grid of the window, and $\mu$ is found by solving
$N_\mathrm{el}(\mu, T) = n_\mathrm{el}$. At
$T = 0$ the Fermi--Dirac distribution is a step function, and solving on the
grid would round $\mu$ to the grid spacing. At $T = 0$, $\mu$ is therefore
the energy at which the number of states below it, counted by the
tetrahedron method, equals $n_\mathrm{el}$. This count is continuous in
energy.

\section{Extracting the expansion coefficients}
\label{sec:extraction}

The minimization of Sec.~\ref{sec:minimization} gives the equilibrium
lattice parameters $\bm{x}(T)$ at the temperatures of the sweep. For a
hexagonal crystal $\bm{x}(T) = (a(T), c(T))$. This section computes the axial thermal expansion coefficients of
Eq.~(\ref{eq:alpha}) from $a(T)$ and $c(T)$. With the SSCHA $\Fvib$ of
Eq.~(\ref{eq:fvib-mean}), a function of temperature, the Einstein fit, is
fitted to $a(T)$ and $c(T)$ independently, and the derivative of
Eq.~(\ref{eq:alpha}) is taken of the fitted functions. With the harmonic
$\Fvib$ of Eq.~(\ref{eq:fharm}), the derivative can be taken of $a(T)$ and $c(T)$ directly by central
differences.

The third law of thermodynamics requires the thermal expansion
coefficients to be zero at $T = 0$, as Appendix~\ref{app:derivative} shows.
Each Einstein term of Eq.~(\ref{eq:einstein}) below has zero slope
at $T = 0$, so the thermal expansion coefficients computed from the fitted
functions are zero there. A general smoothing method does not make the
slope zero at $T = 0$. The fitted functions are also continuous in
temperature, so the thermal expansion coefficients can be computed at any
temperature in the range of the sweep. The fit removes the scatter of $a(T)$ and $c(T)$
from one temperature to the next that the sampling error of the
SSCHA can cause. Sections~\ref{sec:terms} and \ref{sec:seed} measure this
scatter.

\subsection{The Einstein form}

In the Einstein fit each lattice parameter is a sum of Einstein
terms,\cite{chatterji-einstein-gruneisen,angel-eosfit}
\begin{equation}
  c(T) = c_0 + \sum_{m=1}^{M} A_m \frac{\theta_m}{e^{\theta_m/T} - 1},
  \label{eq:einstein}
\end{equation}
where $c_0$ is the value of $c$ at $T = 0$, and each term $m$ has an
amplitude $A_m$ and an Einstein temperature $\theta_m$. The same form is
fitted to $a(T)$. Each term has zero value and zero slope at $T = 0$, so
the model satisfies the third law by construction. Two terms of opposite
sign describe a lattice parameter that contracts at low temperature and
expands at high temperature, as the $c$ axis of $\alpha$-Ti does.
The number of terms $M$ has to be chosen. Section~\ref{sec:terms} compares
$M = 2$ and $M = 3$, and it shows that this choice changes the axial
thermal expansion coefficients more than the scatter that the fit removes.

The expansion coefficients of Eq.~(\ref{eq:alpha}) are computed by
differentiating Eq.~(\ref{eq:einstein}) with respect to temperature and
dividing by $c(T)$, and in the same way for $a(T)$. The derivative of each
Einstein term is computed in closed form,
\begin{equation}
  \frac{d}{dT}\frac{\theta}{e^{\theta/T}-1}
  = \frac{(\theta/2T)^2}{\sinh^2(\theta/2T)},
  \label{eq:einstein-derivative}
\end{equation}
where the right-hand side is written with $\sinh$ because this form is
numerically stable at low temperature.

\subsection{The Einstein terms and the Gr\"uneisen relation}

Equation~(\ref{eq:einstein}) can be interpreted physically as the
Gr\"uneisen relation integrated over
temperature.\cite{wallace-thermodynamics} For the $c$ axis of a hexagonal
crystal the Gr\"uneisen relation, Eq.~(\ref{eq:alpha-c-compliance}) of
Appendix~\ref{app:derivative}, reads
\begin{equation}
  \frac{d\ln c}{dT} = \frac{1}{V} \sum_{\bm{q}\nu}
  \left( 2 s_{ac}\gamma^{\bm{q}\nu}_{a} + s_{cc}\gamma^{\bm{q}\nu}_{c} \right)
  C_{\bm{q}\nu}(T),
  \label{eq:gruneisen}
\end{equation}
where $C_{\bm{q}\nu}$ is the heat capacity of the mode $\bm{q}\nu$,
$\gamma^{\bm{q}\nu}_{a}$ and $\gamma^{\bm{q}\nu}_{c}$ are its Gr\"uneisen
parameters for strains along $a$ and $c$, $s_{ac}$ and $s_{cc}$ are elastic
compliances, and $V$ is the volume. The factor in parentheses is the weight
of each mode. It contains the Gr\"uneisen parameters for both axes, because
$a$ and $c$ are coupled through $s_{ac}$. Integrating
Eq.~(\ref{eq:gruneisen}) from $T = 0$, with the weights taken as constant, replaces the heat capacity of
each mode by its thermal energy,
$\hbar\omega_{\bm{q}\nu} / (e^{\hbar\omega_{\bm{q}\nu}/\kB T} - 1)$. The
change of $c$ is small, so $c(T) - c_0$ is proportional to
$\ln c(T) - \ln c_0$ to first order. Equation~(\ref{eq:einstein}) is the
integrated sum with the modes gathered into $M$ groups. Each group is
represented by one Einstein temperature $\theta_m$, and its amplitude $A_m$
contains the sum of the weights of the modes in the group.

The sign of $A_m$ is the sign of the summed weights of its group. A
negative $A_m$ represents a group of modes whose excitation shortens the
axis. The weight contains $2 s_{ac}\gamma^{\bm{q}\nu}_{a}$, so a group of
modes can shorten $c$ even when their Gr\"uneisen parameters are all
positive. Amplitudes of mixed sign describe an axis that contracts while
only the groups with negative amplitudes are excited, and expands once the
other groups are excited too. The third law is satisfied because the
thermal energy, which contains no zero-point part, vanishes at $T = 0$ with
zero slope. A few terms suffice because the model only has to reproduce
the coarse dependence of the weights on frequency, not the modes one by
one.

\subsection{The least-squares fit}

Each fit is an unweighted nonlinear least-squares fit made with
\texttt{scipy.optimize.curve\_fit} of SciPy.\cite{scipy} With no bounds on
the parameters, \texttt{curve\_fit} uses the Levenberg--Marquardt
algorithm.\cite{levenberg,marquardt} The fit is run from many starting
parameters. The starting Einstein temperatures are taken from a list of
trial values, and the starting amplitudes are given various patterns of
signs.

Every fit that converges is checked. A fit is rejected when one of
its Einstein temperatures is not positive. A least-squares fit can also
converge to a curve with the wrong shape at low temperature. The $c$ axis
of $\alpha$-Ti becomes shorter with increasing temperature at low
temperature and longer at high temperature. The decrease is small compared
with the increase, and it contributes little to the sum of squared
residuals. A fit can therefore miss the decrease, or give a decrease much
steeper than that of $c(T)$. Phonopy rejects such fits with a heuristic
check on the slope of the fitted curve, so that a negative thermal expansion such as that of the $c$ axis
of $\alpha$-Ti is fitted well. The details of the check are in the source
code of phonopy. From the fits that remain, phonopy uses the fit with the smallest sum of
squared residuals. When every fit is rejected, phonopy stops with an error that says
that no acceptable Einstein fit was found.

\section{Procedure}
\label{sec:recipe}

\begin{figure*}
  \centering
  \input{figures/workflow}
  \caption{\label{fig:workflow}%
    The procedure. Solid boxes are commands, dashed boxes are calls to the
    electronic-structure calculator, and rounded boxes are the files they
    write. The left column produces a dataset from which harmonic force
    constants are computed. When this dataset alone is analysed, the
    calculation uses harmonic force constants, as in the quasi-harmonic
    approximation. The right column replaces the vibrational free energy
    with one from the SSCHA, which uses a machine-learning potential fitted
    at each strained cell, and passes it to the same analysis.}
\end{figure*}
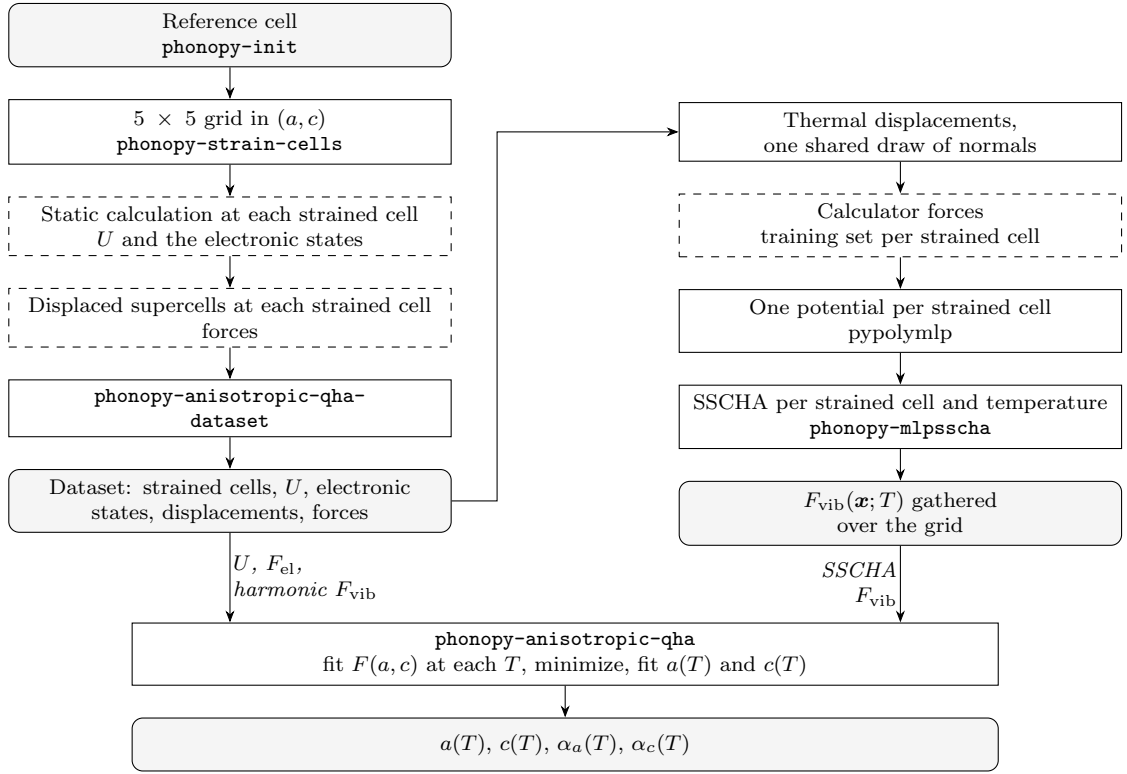

The procedure needs one unit cell of the crystal as its input. This unit
cell may be relaxed with the static energy, which places the fitting region
near the minimum of $U$. The command \texttt{phonopy-symmetry} writes the
standardized conventional unit cell of a given structure. The command
\texttt{phonopy-init} saves this unit cell as the reference cell, together
with a choice of supercell size.
The command
\texttt{phonopy-strain-cells} makes the strained cells from the reference
cell at chosen values of the lattice parameters. The calculator then runs a
static calculation on each strained cell, which gives $U$ and the electronic
states, and computes the forces of supercells with displacements, from
which the harmonic force constants of that strained cell are obtained. The command
\texttt{phonopy-anisotropic-qha-dataset} collects these calculations into
one file. The command \texttt{phonopy-anisotropic-qha} reads that file, fits
the free-energy surface at each temperature and minimizes it.
The left column of Fig.~\ref{fig:workflow} shows these commands, the calls
to the calculator and the files they write. The right column of the figure is
the variant of Sec.~\ref{sec:tdfc}, in which the harmonic vibrational free
energy is replaced by one computed with temperature-dependent force
constants.

The implementation reads the static internal energy and the electronic
eigenvalues from the output of VASP. At the time of writing the procedure
works only with VASP.

\subsection{The reference cell}

The reference cell is a standardized conventional unit cell of the crystal.
It may be relaxed with the static energy. The command \texttt{phonopy-init}
saves it together with the supercell matrix, the primitive matrix and the
name of the calculator, and every later command reads what it saved. The
supercell matrix and the primitive matrix are the transformation matrices
that represent the basis vectors of the supercell and of the primitive cell
with respect to the basis vectors of the reference cell. The symmetry of
this unit cell determines the free lattice parameters, and the strained cells are
built from it by changing the lattice parameters. The energies of
Eq.~(\ref{eq:total}) are given per primitive cell, which the saved primitive
matrix defines.

The relaxation of the reference cell does not have to be tight. The lattice
parameters of the reference cell are a guide for choosing the values
$\bm{x}_j$ at which the strained cells are made. The polynomial of
Eq.~(\ref{eq:poly}) is fitted in those values, centred on their mean, so the
lattice parameters of the reference cell do not enter the fit. The fitting region has to contain the equilibrium
lattice parameters $\bm{x}(T)$ at every temperature that is analysed, because
outside the fitting region the fitted surface is an extrapolation. The ranges
given to \texttt{phonopy-strain-cells} decide the fitting region. The
minimum of $U$ is a convenient centre for that region, and $\bm{x}(T)$ is not
at that minimum, because the vibrational and the electronic free energy move
it (Sec.~\ref{sec:results}).

The space group that phonopy finds in this unit cell
determines which of the lattice parameters are free.
Unless the unit cell is the conventional unit cell, the command
\texttt{phonopy-strain-cells} stops and prints an error message. In that case
the command \texttt{phonopy-symmetry} converts the structure into the
conventional unit cell, which is the unit cell the procedure needs.

\subsection{Sampling strained cells under the crystal symmetry}
\label{sec:free-parameters}

The components of $\bm{x}$ are the free lattice parameters, the lattice
parameters that can change independently under the symmetry of the
crystal. A cubic crystal has one free lattice parameter. A hexagonal,
tetragonal or rhombohedral crystal has two, and an orthorhombic crystal has
three. The cell angles are held fixed. In a monoclinic or triclinic crystal
at least one angle is free, so the procedure does not apply to these
crystals.

The unit cell at each value $\bm{x}_j$ of Sec.~\ref{sec:minimization} is made
by straining the reference cell. The reference cell is a conventional unit
cell, which is used because the symmetry determines its shape uniquely. Its
three basis vectors are written $\mathbf{a}$, $\mathbf{b}$ and $\mathbf{c}$.
A strained cell is made by changing the lengths of these vectors, and basis
vectors that the symmetry relates to each other are changed by the same
ratio. The symmetry of a hexagonal crystal, for example, requires
$\mathbf{a}$ and $\mathbf{b}$ to have the same length, $b = a$, so the free
lattice parameter $a$ is the length of both vectors and the free lattice
parameter $c$ is the length of $\mathbf{c}$. For a rhombohedral crystal the
hexagonal conventional unit cell is used.

\subsection{The static calculations}

The analysis accepts any set $\{\bm{x}_j\}$, as long as the number of
strained cells $N_x$ is at least the number of coefficients of the polynomial
of Eq.~(\ref{eq:poly}). A regular grid over the free lattice parameters is
recommended, for example $4 \times 4$ or $5 \times 5$ when two lattice
parameters are free, and the command \texttt{phonopy-strain-cells} makes such
a grid. The user gives the range and the number of values for each free
lattice parameter. In this report the grid is $5 \times 5$ over $(a, c)$ for
both phases, and every result that follows comes from the strained cells of
that grid. Table~\ref{tab:settings} gives the grid chosen for each phase.
Each strained cell is
made from the reference cell by changing the lengths of its basis vectors, as
described in Sec.~\ref{sec:free-parameters}, and the fractional coordinates
of the atoms are kept.

Each strained cell is relaxed in its internal coordinates if the crystal
has free internal coordinates. In $\alpha$- and $\omega$-Ti the symmetry
fixes the positions of the atoms inside the unit cell, so no relaxation is
needed. A static single-point calculation is then run on each strained cell.
The Brillouin zone is sampled on a regular mesh that is compatible with the
crystallographic point group. The tetrahedron method needs such a mesh,
because the implementation maps the $k$ points onto the irreducible part of
the mesh with the operations of that point group. The
calculator is asked to write the eigenvalues at every $k$ point together
with the mesh, so that the analysis can later compute $\Fel$ from them with
the tetrahedron method. The mesh is given as the number of divisions along
each axis, and the same numbers are used at every strained cell. Appendix~\ref{app:settings} describes the
settings that a calculator derives from the unit cell, what happens when they
come out differently at different strained cells, and how to check for it.

\subsection{The harmonic phonon calculations}

The harmonic force constants of each strained cell are computed from the
forces of supercells with displacements. These supercells are built on the
strained cell after its internal relaxation, and their forces are computed
with the calculator. The supercell matrix and the displacement distance are
the same at every strained cell.

The forces at a finite displacement are not purely harmonic. The force
constants computed from them therefore have an error, and the error becomes
smaller as the displacement distance is reduced. A distance that is too small
amplifies the noise of the calculator instead, because the force constants
are the forces divided by the displacement.

The user chooses the size of the supercell. Computing the axial thermal
expansion coefficients with supercells of increasing size and comparing them
is recommended. In a metal the $k$-point mesh and the smearing of the
supercell calculations have to be tested together with the size of the
supercell. A larger supercell needs fewer divisions of the mesh for the same
sampling of the Brillouin zone, and the smearing width that the calculation
needs depends on how dense that sampling is. Testing the three together is
expensive, so the choice is made in part from experience.
Section~\ref{sec:details} gives the supercell, the mesh and the smearing used
here.

\subsection{The dataset}

The command \texttt{phonopy-anisotropic-qha-dataset} collects the static and
supercell calculations of all strained cells into one file, and the analysis at
the following step reads that file. For each strained cell the file holds
\begin{itemize}
  \item the strained cell after its internal relaxation;
  \item the supercell matrix and the primitive matrix;
  \item the displacements and the forces of the supercells;
  \item the static internal energy;
  \item the electronic states.
\end{itemize}
The file holds the displacements and the forces of the supercells, and not
the force constants. The force constants are computed when the file is
analysed.

The static calculations and the supercell calculations are given to
\texttt{phonopy-anisotropic-qha-dataset} as two lists, and they are paired by
their position in those lists. The user runs the static and supercell
calculations of a strained cell separately, and a mistake in the two lists is
easy to make. The lists can be given in different orders, or a strained cell
can be left out of one of them. The command therefore checks every pair, and
stops and prints an error message when the two calculations of a pair do not
belong to the same strained cell.

\subsection{The analysis}

The command \texttt{phonopy-anisotropic-qha} carries out the analysis. The
force constants of every strained cell are computed from the stored
displacements and forces. The three terms of Eq.~(\ref{eq:total}) are then
evaluated at each temperature that the user gives, and the surface is fitted
and minimized at each temperature. The command writes the equilibrium lattice
parameters, the axial thermal expansion coefficients and the free-energy
contour maps.

The harmonic vibrational free energy of Eq.~(\ref{eq:fharm}) is computed from
the force constants, and the phonons are sampled on a $\bm{q}$-point mesh for
it. At this step the mesh is given as a length $l$ in \AA, and a larger $l$
gives a denser mesh. The user can give a different value from the default,
$l = 200$~\AA. When the vibrational free
energy comes from the variant of Sec.~\ref{sec:tdfc}, the analysis reads it
from that calculation, and the mesh of this step is not used.

\subsection{Temperature-dependent force constants}
\label{sec:tdfc}

The displacements used for the harmonic force constants are small, and the
forces they give are almost harmonic. Anharmonic effects appear at the
larger displacements of thermal motion. The SSCHA needs the forces of
thousands of supercells with displacements of that size at every strained
cell and every temperature, and computing them with the calculator is too
expensive. A
machine-learning potential is therefore fitted at each strained cell, and
the SSCHA iterations of Eq.~(\ref{eq:fiteration}) are run with it. Here the
potentials are polynomial machine-learning potentials (polynomial MLPs)
fitted with pypolymlp.\cite{pypolymlp} The MLP computes the energies and the
forces of the supercells that each iteration draws. The energies enter the
average of Eq.~(\ref{eq:sscha-estimate}), and the force constants of the
next iteration are fitted to the forces. The accuracy of these energies and
forces follows the settings of the calculations that the MLP was trained on.
Those calculations are made on supercells, and the static calculation of a
strained cell is made on the unit cell. Raising the plane wave cutoff and the
$k$-point density costs far less on the unit cell, so $U$ can be computed
more accurately than the energies the MLP was fitted to.

The static internal energy $U$ is taken from the calculator here, and not
from the MLP. The analysis can take $U$ from the MLP instead. The minimum of
Eq.~(\ref{eq:total}) is at the lattice parameters where the
gradient of $U$ cancels the gradients of $\Fvib$ and $\Fel$, as
Sec.~\ref{sec:results} shows. A small error in the gradient of $U$
therefore moves the minimum, and the axial thermal expansion coefficients
change with it.

Each MLP is evaluated only at its own strained cell, where it has to
reproduce the forces and the energies of supercells with displacements. The
dependence of $\Fvib$ on the lattice parameters comes from the SSCHA runs
at the different strained cells, each with its own MLP, and from the
surface fit over them.

\subsubsection{The training structures}
\label{sec:training}

The MLP of each strained cell is fitted to the energies and forces
that the calculator computes for a set of supercells with displacements.
These are
called the training supercells. Atoms are displaced according to the
thermal distribution of the harmonic crystal defined by the harmonic force
constants of that strained cell. At each $\bm{q}$ point commensurate with
the supercell, the probability distribution function along each normal
coordinate is Gaussian with variance
\begin{equation}
  \sigma_{\bm{q}\nu}^2(T) = \frac{\hbar}{2\omega_{\bm{q}\nu}}
  \coth \frac{\hbar\omega_{\bm{q}\nu}}{2\kB T}
  = \frac{\hbar}{2\omega_{\bm{q}\nu}}
    \left[ 1 + 2 n_{\bm{q}\nu}(T) \right],
  \label{eq:sigma}
\end{equation}
where
\begin{equation}
  n_{\bm{q}\nu}(T) =
  \frac{1}{e^{\hbar\omega_{\bm{q}\nu}/\kB T} - 1}
  \label{eq:bose}
\end{equation}
is the Bose--Einstein distribution. The second form of
Eq.~(\ref{eq:sigma}) separates the variance into two parts. The first term
is the zero-point motion, and it is present at every temperature, including
$T = 0$. The second term is the thermal
occupation.

One supercell is drawn by taking standard normals $\xi_{\bm{q}\nu}$, one for
each normal coordinate, and transforming them to Cartesian
displacements.\cite{phonopy-implementation} The distribution is computed
once from the harmonic force constants, and it is not made self-consistent
with the temperature-dependent force constants of the SSCHA.

The sampling matrix at one $\bm{q}$ point is
\begin{equation}
  A_{\bm{q}} = W_{\bm{q}} \operatorname{diag}
  \left( \sigma_{\bm{q}1}, \ldots, \sigma_{\bm{q}3N} \right)
  W_{\bm{q}}^{\dagger},
  \label{eq:sampling-matrix}
\end{equation}
where $W_{\bm{q}}$ is the matrix of eigenvectors of the dynamical matrix at
that $\bm{q}$ point,
\begin{equation}
  D_{\bm{q}} = W_{\bm{q}} \operatorname{diag}
  \left( \omega_{\bm{q}1}^2, \ldots, \omega_{\bm{q}3N}^2 \right)
  W_{\bm{q}}^{\dagger}.
  \label{eq:dynamical-matrix}
\end{equation}
The drawn amplitudes at that $\bm{q}$ point are $A_{\bm{q}}\xi$, where
$\xi$ is the vector drawn from the standard normals $\xi_{\bm{q}\nu}$.
Reference~\onlinecite{phonopy-implementation} uses
$W_{\bm{q}} \operatorname{diag}(\sigma)\xi$ instead, which is
$A_{\bm{q}}\xi$ with the unitary factor $W_{\bm{q}}^{\dagger}$ removed. Multiplying the standard normals by that unitary factor gives
another set of standard normals, $\xi' = W_{\bm{q}}^{\dagger}\xi$, which is
also valid. Equation~(\ref{eq:sampling-matrix}) then gives
$A_{\bm{q}}\xi = W_{\bm{q}} \operatorname{diag}(\sigma)\,\xi'$. The same $\xi$
gives $A_{\bm{q}}\xi$ with one form and
$W_{\bm{q}} \operatorname{diag}(\sigma)\xi$ with the other, and these are
different vectors. Equations~(\ref{eq:sampling-matrix})
and~(\ref{eq:dynamical-matrix}) differ only in the diagonal factor, and
$\sigma_{\bm{q}\nu}$ depends only on $\omega_{\bm{q}\nu}$. Therefore a
rotation of the eigenvectors within a degenerate subspace does not change
$A_{\bm{q}}$.

The MLPs are fitted independently at each strained cell, so their
errors vary from one strained cell to the next. The surface fit of
Eq.~(\ref{eq:poly}) cannot distinguish this variation from a real
dependence of $\Fvib$ on the lattice parameters. To make the variation
smooth, $\xi$ is drawn once and turned into displacements at every strained
cell with the matrices $A_{\bm{q}}$ of that strained cell. The displacement
fields of neighbouring strained cells are then similar, and the errors of
the MLPs fitted to them can vary smoothly from one strained cell to
the next.

The similarity was measured as a correlation coefficient. The displacement
field of each strained cell was correlated with the displacement field of
the central strained cell, over every supercell, atom and Cartesian component
of one draw at 250~K. The coefficient is $0.9989$ to $0.99997$ across the
25 strained cells of $\alpha$-Ti and $0.9995$ to $0.99999$ for $\omega$-Ti.
When each strained cell draws its own normals instead, the coefficient is
$-0.026$ to $+0.023$ in $\alpha$-Ti and $-0.031$ to $+0.021$ in $\omega$-Ti,
as expected for independent fields. The shared draw gives similar fields
only with $A_{\bm{q}}$ of Eq.~(\ref{eq:sampling-matrix}). When the amplitudes
are formed with $W_{\bm{q}} \operatorname{diag}(\sigma)$ instead, the
correlation decreases to between $0.13$ and $0.47$ in $\alpha$-Ti and between
$-0.07$ and $0.44$ in $\omega$-Ti.

The training temperatures cover the temperatures that the SSCHA runs use,
because an MLP is less accurate outside the range of its training
data. The same number of supercells is drawn at each training temperature,
and the sets of one strained cell are merged into one list that takes one
structure from each temperature in turn. The first $n$ structures of the
merged list therefore hold the training temperatures in nearly equal parts
for any $n$. A smaller training set can then be taken as the first $n$
structures of the list, which is how the training sets of
Sec.~\ref{sec:ladder} were made.

\subsubsection{Fitting the MLPs}

One MLP is fitted per strained cell, to that strained cell's own
training set. A polynomial MLP is a polynomial in
invariants that describe the environment of each atom. It is linear in its
coefficients, which are fitted by ridge regression. The model parameters of
the MLP determine which invariants and which of their products enter the
polynomial, and with them the number of coefficients. The model parameters
and the ridge penalty of the regression have to be set to the same values at
every strained cell. The ridge penalty needs care,
because it is often chosen automatically, by the smallest test error, and
that choice gives different values at some neighbouring strained cells. Each
change would put a step into $\Fvib(\bm{x})$, which the surface fit cannot
distinguish from a real dependence on the lattice parameters. The choice of
model parameters changes both the accuracy of the MLP and the cost of the
evaluation, and Sec.~\ref{sec:ladder} measures both.

\subsubsection{The SSCHA runs}

The command \texttt{phonopy-mlpsscha} carries out one run. A run computes
the free energy at one strained cell and one temperature with that strained
cell's MLP. The temperatures are chosen
before the first run, and every strained cell uses the same temperatures,
because the surface fit at each temperature needs $\Fvib$ of every strained
cell at that temperature. The runs at different strained cells and
temperatures are independent of each other, so they are submitted as
separate jobs on a cluster. Every run of a sweep uses the same random seed.
Section~\ref{sec:seed} measures how this choice changes the axial thermal
expansion coefficients.

Each run stores the free energy $F^{(k)}$ of every iteration. The mean of
Eq.~(\ref{eq:fvib-mean}) is taken afterwards, and the length of the
transient is chosen from the stored values. Changing the length of the
transient therefore needs only a new mean of the stored values.

\section{Computational details}
\label{sec:details}

\subsection{Crystal structures}

$\alpha$-Ti is hexagonal close packed, space group $P6_3/mmc$ (No.~194),
with two atoms in the conventional unit cell on the Wyckoff position $2c$.
$\omega$-Ti is space group $P6/mmm$ (No.~191), with three atoms on the
Wyckoff positions $1b$ and $2c$. In both structures every atom occupies a
fixed Wyckoff position, so there is no internal coordinate to relax. Each
strained cell therefore needs no internal relaxation, and the supercells
with displacements are built directly on the strained cells. Both crystals are hexagonal, so the free
lattice parameters are $a$ and $c$ in both.

\begin{figure*}
  \includegraphics{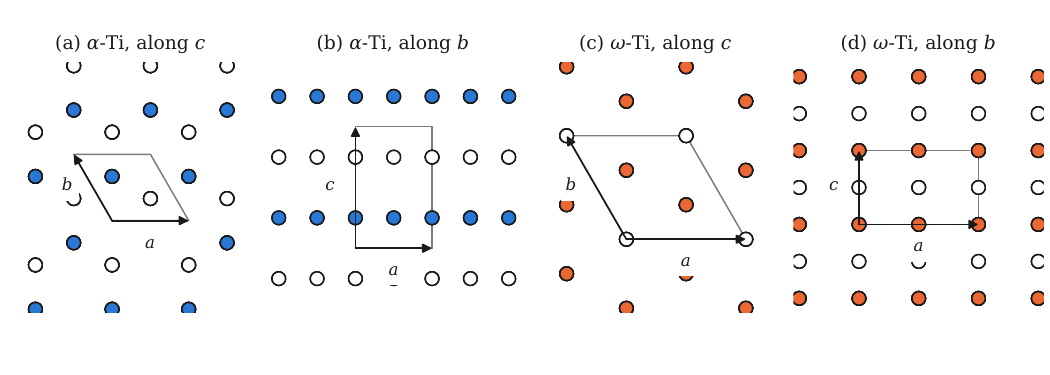}
  \caption{\label{fig:structure}%
    The two structures, drawn from the central strained cell of each phase: (a)
    and (b) $\alpha$-Ti, (c) and (d) $\omega$-Ti, seen down the $c$ axis and
    along the $b$ axis. Filled circles are the lower layer of the unit cell and
    open circles the upper one. All four panels are drawn at the same scale.}
\end{figure*}

Figure~\ref{fig:structure} shows the two structures. The close-packed layers of $\alpha$-Ti are stacked ABAB, and each
layer is offset from the one below. In $\omega$-Ti a flat honeycomb layer on
$2c$ alternates with a simple hexagonal layer on $1b$. The $c$
axis of $\omega$-Ti is shorter, $2.83$~\AA{} against $4.66$~\AA{} for
$\alpha$-Ti, and it covers two inequivalent layers rather than two
equivalent ones.

\subsection{First-principles calculations}

All calculations use the projector augmented-wave (PAW)
method\cite{PAW-Blochl-1994} as implemented in
VASP,\cite{VASP-Kresse-1995,VASP-Kresse-1996,VASP-Kresse-1999} with the
Perdew--Burke--Ernzerhof exchange-correlation
functional.\cite{Perdew-PBE-1996} The PAW potential for Ti treats the $3p$,
$3d$ and $4s$ electrons as valence.

The static calculations that give $U$ and the electronic states use a plane
wave cutoff of 500~eV, an energy convergence of $10^{-8}$~eV, and the
tetrahedron method with Bl\"ochl corrections for the Brillouin-zone
integration. The $k$-point mesh is $\Gamma$ centred, given explicitly, and
the same for all 25 strained cells. The eigenvalues that $\Fel$ is
computed from are taken from a second, denser mesh. The eigenvalues on the
denser mesh are computed non-self-consistently from the charge density
obtained on the first mesh. Table~\ref{tab:settings} gives the two meshes of
each phase.

The supercell calculations that give the training forces use a plane wave
cutoff of 300~eV and a first-order Methfessel--Paxton smearing of 0.2~eV,
with the same energy convergence as the static calculations. They also use
the support grid for the augmentation charges (\texttt{ADDGRID = .TRUE.}).
In the author's experience this setting matters for the accuracy of the
fitted MLP.

The static and the supercell calculations do not share their settings, so
Eq.~(\ref{eq:total}) sums terms computed on two slightly different
Born--Oppenheimer surfaces. The training calculations use the lower cutoff
because a cutoff of 500~eV would increase the cost of every training
structure. Even with
the same settings a difference would remain, because the SSCHA iterations
use the energies of the fitted potential and not those of the calculator.
This difference is not corrected, and it is not measured in this report.

\subsection{Force constants and supercells}

The harmonic force constants are fitted by symfc\cite{symfc} to supercells
with displacements built on each strained cell, with a displacement distance
of 0.03~\AA. One systematic displacement per
strained cell is sufficient for $\alpha$-Ti, and two are needed for
$\omega$-Ti.

The supercell is $4\times4\times4$ of the two-atom unit cell for $\alpha$-Ti,
which is 128 atoms, and $3\times3\times4$ of the three-atom unit cell for
$\omega$-Ti, which is 108 atoms. The $k$-point mesh of the supercell
calculations is $4\times4\times2$ for $\alpha$-Ti and $4\times4\times4$ for
$\omega$-Ti. Everything else in the
calculations of the two phases is the same, including the training temperatures, the number
of training structures, the model parameters of the MLP, the ridge penalty and every
parameter of the SSCHA runs.

\subsection{MLPs and SSCHA runs}
\label{sec:mlp}

The training set of a strained cell consists of 50 supercells drawn at each
of 0, 100, 250 and 400~K. The four sets are merged into one list of 200
supercells, as described in Sec.~\ref{sec:training}. Of the 200 supercells,
180 are used for the fit and 20 are held out as a test set. The model
parameters are set to the default values of phonopy's pypolymlp interface.
The invariants are of the gtinv type, with \texttt{gtinv\_order}~3 and
\texttt{gtinv\_maxl}~$(8, 8)$, and eleven Gaussian pair parameters inside a
cutoff of 8~\AA. The polynomial is of \texttt{model\_type}~3 with
\texttt{max\_p}~2. With these values the MLP has 781 coefficients. The
ridge penalty is fixed at $10^{-3}$ at every strained cell.

Each SSCHA run draws $N = 2500$ supercells per iteration and runs 11
iterations. The first iteration is the transient, so $k_0 = 1$, and the
remaining ten are averaged by Eq.~(\ref{eq:fvib-mean}), so $K = 10$. The
$\bm{q}$-point mesh is $l = 400$~\AA, and every run uses the same random
seed, 1000. One run is made for each of the 25 strained cells and each
of the 41 temperatures from 0 to 400~K in steps of 10~K, in both phases.
This sweep gives the results of Sec.~\ref{sec:results}, and it is called
the reported sweep.

\subsection{The analysis settings}

The analysis fits the free energy with the polynomial of
Eq.~(\ref{eq:poly}), of total degree 3 in $a$ and $c$. The polynomial has
ten coefficients and is fitted to the 25 strained cells. The Einstein fit of
$a(T)$ and $c(T)$ has three Einstein terms, $M = 3$, and it is made to all
41 temperatures of the sweep, from 0 to 400~K. The coefficients are reported
up to 300~K.

The calculation with harmonic force constants, which uses the $\Fvib$ of
Eq.~(\ref{eq:fharm}), is called the harmonic route in this report. The
calculation with temperature-dependent force constants from the SSCHA is
called the SSCHA route. On the harmonic route the analysis computes the
phonon frequencies itself, with the $\bm{q}$-point mesh given as the length
$l = 400$~\AA. That mesh is the same as in the SSCHA runs.

\begin{table}
  \caption{\label{tab:settings}%
    Settings of the two calculations. The sampled ranges are the lattice
    parameters of the conventional unit cell at the corners of the
    $5 \times 5$ grid.}
  \begin{ruledtabular}
    \begin{tabular}{lcc}
      & $\alpha$-Ti & $\omega$-Ti \\
      \hline
      Space group & $P6_3/mmc$ (194) & $P6/mmm$ (191) \\
      Atoms in the unit cell & 2 & 3 \\
      Sampled $a$ (\AA) & 2.9200--2.9680 & 4.5600--4.6180 \\
      Sampled $c$ (\AA) & 4.6280--4.7020 & 2.8180--2.8580 \\
      Supercell & $4\times4\times4$ & $3\times3\times4$ \\
      Atoms in the supercell & 128 & 108 \\
      Static $k$ mesh & $34\times34\times18$ & $27\times27\times38$ \\
      $k$ mesh for $\Fel$ & $68\times68\times36$ & $54\times54\times76$ \\
      Irreducible $k$ points & 7980 & 10569 \\
      Supercell $k$ mesh & $4\times4\times2$ & $4\times4\times4$ \\
    \end{tabular}
  \end{ruledtabular}
\end{table}

\section{Results}
\label{sec:results}

This section gives the results for $\alpha$- and $\omega$-Ti on the
harmonic route and on the SSCHA route. The results of the SSCHA route are
those of the reported sweep.

\begin{figure*}
  \includegraphics{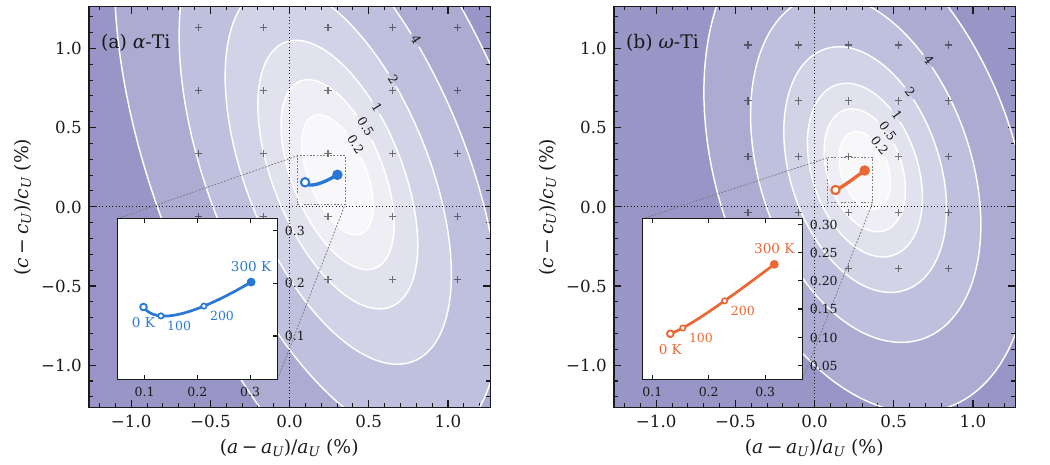}
  \caption{\label{fig:surface}%
    The fitted free-energy surface at 300~K and the path of its minimum from
    0 to 300~K, for (a) $\alpha$-Ti and (b) $\omega$-Ti. Both panels are
    drawn against the strain of each phase measured from $a_U$ and $c_U$, the
    lattice parameters at the minimum of its static energy $U$. The shading
    shows bands between the contour levels $0.2$, $0.5$, $1$, $2$, $4$ and
    $8$~meV above the minimum per primitive cell, from light to dark, with
    the same levels in both panels. Crosses are the 25 strained cells. The
    path of the minimum covers about a fifth of the fitting region, and each
    panel shows the path magnified in an inset. Open circles mark 0, 100 and
    200~K, and a filled circle marks 300~K. The 0~K point is not at the
    origin, and the offset is the zero-point expansion.}
\end{figure*}

\subsection{Free-energy surfaces and the path of the minimum}

Figure~\ref{fig:surface} shows the fitted $F(a, c)$ of each phase at
300~K, and the path of the minimum from 0 to 300~K drawn over it. Both
panels use the same axes and the same contour levels, and the crosses are
the 25 strained cells.

The strains $\epsilon_a$ and $\epsilon_c$ of $a$ and $c$ are measured from
$a_U$ and $c_U$, the lattice parameters that minimize the static energy $U$
alone, rather than from the equilibrium lattice parameters at $T = 0$. The static minimum is a
property of the electronic-structure calculation alone, while the lattice
at $T = 0$ already includes the zero-point term and so depends on the
vibrational free energy that is used. The 0~K point of each path is
therefore not at the origin.

Near its minimum the free-energy surface is a valley. In the strains the
valleys of the two phases are similar. At the 300~K minimum the curvature
per atom $\partial^2 F/\partial \epsilon_a^2$ is $54.1$~eV for $\alpha$-Ti
and $56.3$~eV for $\omega$-Ti, and $\partial^2 F/\partial \epsilon_c^2$ is
$19.2$ and $24.7$~eV. A strain $\epsilon_a$ changes $b$ with $a$, because
$b = a$ in both phases, while a strain $\epsilon_c$ changes $c$ alone. Two
axes are strained in the first case and one in the second, which is part of
why the curvature along $a$ is the larger. The cross term
$\partial^2 F/\partial \epsilon_a \partial \epsilon_c$ is $+16.8$ and
$+11.8$~eV. The cross term is comparable to the diagonal terms, so the
position of the minimum along one axis depends on the position along the
other, and the minimization has to be carried out in both variables at
once.

The same curvatures expressed in the lattice parameters themselves, in eV
per square angstrom and per primitive cell, say how far an error in the free
energy moves the minimum, because $F$ of Eq.~(\ref{eq:total}) is given per
primitive cell and the minimization varies $a$ and $c$. The curvature of
$\alpha$-Ti is $12.48$~eV~\AA$^{-2}$ along $a$ and $1.77$ along $c$. The
curvature of $\omega$-Ti is $8.01$ along $a$ and $9.22$ along $c$. These
units mix the length of the axis with the number of atoms in the primitive
cell, so the strain curvatures above are the ones that compare the two
crystals. The curvature along the $c$ axis of $\alpha$-Ti is the smallest of
these values, partly because the axis is long. A change of $c$ by one
milliangstrom is a smaller strain of a $4.66$~\AA{} axis than of a
$2.83$~\AA{} axis, and the free energy depends on the strain. The cross term
turns the direction in which the surface is softest away from the $c$ axis.
In these units the difference between the curvatures along $a$ and along $c$
is several times the cross term, so that direction stays close to the $c$
axis. An error of a
few microelectronvolts in the free energy therefore displaces $c$ of
$\alpha$-Ti much more than it displaces any other lattice parameter in
either phase. For this reason every setting in
Sec.~\ref{sec:choices} is measured on $\alpha_c$ of $\alpha$-Ti.

\begin{figure*}
  \includegraphics{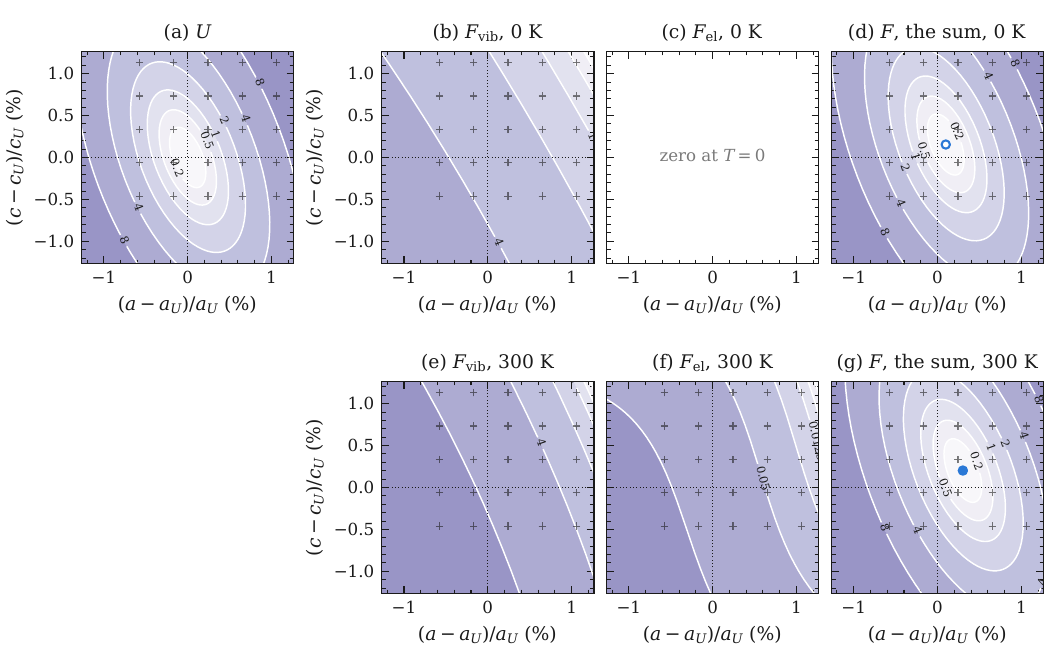}
  \caption{\label{fig:decomposition}%
    The free-energy surface of $\alpha$-Ti split into the three terms of
    Eq.~(\ref{eq:total}): (a) the static energy, which does not depend on
    temperature; (b), (e) the vibrational free energy; (c), (f) the electronic
    free energy; and (d), (g) their sum. The upper row is at 0~K and the
    lower row at 300~K. At 0~K the vibrational
    free energy is the zero-point term alone, and the electronic free energy
    is zero. Each panel shows the energy measured from its own minimum, over the same fitting region and
    in the same strain coordinates as Fig.~\ref{fig:surface}. The panels of
    $U$, $\Fvib$ and the sum use the contour levels of that figure, $0.2$ to
    $8$~meV. Panel (f) is a hundred times smaller and uses $0.0025$ to
    $0.1$~meV. Crosses are the 25 strained cells each panel is fitted to. The
    open circle in (d) and the filled circle in (g) mark the minimum of the
    sum, open at 0~K and filled at 300~K as in Fig.~\ref{fig:surface}, in
    the colour of this phase in that figure. The panels of $\Fvib$
    and $\Fel$ tilt the valley of $U$.
    \texttt{phonopy-anisotropic-qha} writes the 300~K row of this figure with
    \texttt{-{}-decompose-contours}.}
\end{figure*}

Figure~\ref{fig:decomposition} splits the surface into the three terms of
Eq.~(\ref{eq:total}). The static energy $U$ does not depend on temperature
and is drawn once, in panel (a). The other terms and their sum are drawn at
0~K in the upper row and at 300~K in the lower row. The shape of the valley
comes from the static energy $U$. The
curvatures of $U$ at the 300~K minimum are $12.76$~eV~\AA$^{-2}$ along $a$
and $1.85$ along $c$, with a cross term
$\partial^2 U/\partial a \partial c$ of $2.39$. The corresponding values of
the total are $12.48$, $1.77$ and $2.45$. The vibrational and electronic terms
change the curvature of the surface very little. $\Fvib$ contributes
$-0.28$ and $-0.09$ to the two curvatures, a few per cent of the curvatures
of $U$ and of the opposite sign. The values and the slopes of $\Fvib$, however,
are not small. Over the fitting region $\Fvib$ varies by $9.4$~meV and $U$ by
$12.4$~meV. At the minimum of the total free energy at 300~K the
gradient of $\Fvib$ has a size of $141$~meV~\AA$^{-1}$, and the gradient of
$U$ has a size of $143$~meV~\AA$^{-1}$ and points in nearly the opposite
direction. The gradient of $\Fel$ is small and makes up the difference, and
the three gradients sum to zero at that point.

\begin{figure*}
  \includegraphics{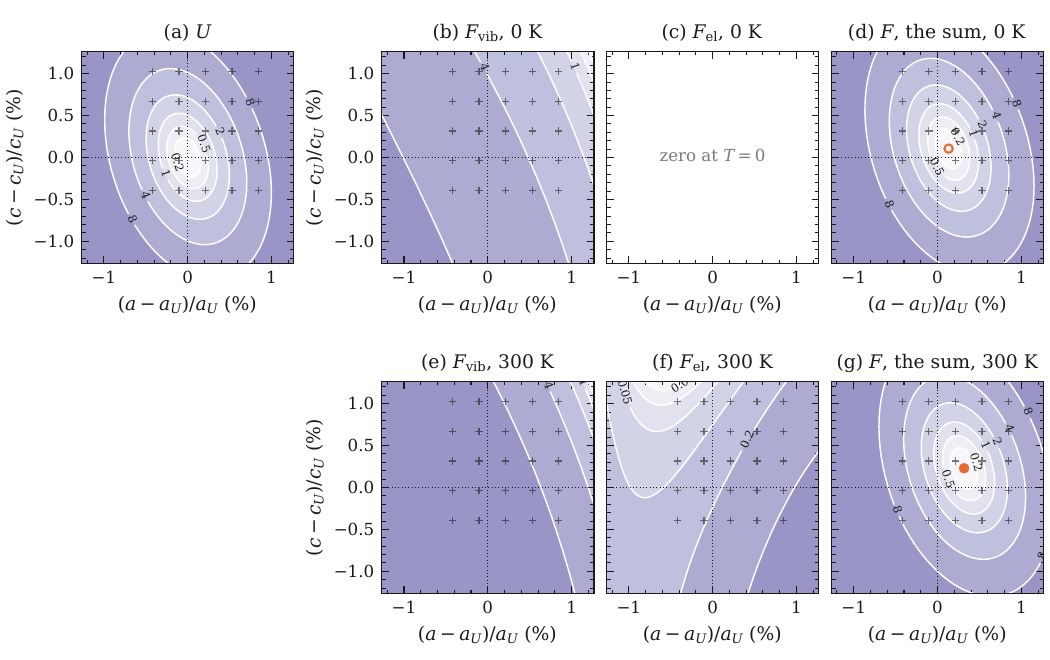}
  \caption{\label{fig:decomposition-omega}%
    The same decomposition for $\omega$-Ti, drawn in the same way as
    Fig.~\ref{fig:decomposition}: (a) the static energy; (b), (e) the
    vibrational free energy; (c), (f) the electronic free energy; and (d), (g)
    their sum, at 0~K in the upper row and at 300~K in the lower row, each
    measured from its own minimum and in the strain coordinates of
    Fig.~\ref{fig:surface}(b), over the same range as
    Fig.~\ref{fig:decomposition}. The panels of $U$, $\Fvib$ and the sum use the
    same contour levels as there, $0.2$ to $8$~meV. Panel (f) uses $0.01$ to
    $0.4$~meV, four times the range of its counterpart in
    Fig.~\ref{fig:decomposition}. $\Fel$ is the term whose size differs most
    between the phases. The valley of $U$ is the round one of $\omega$-Ti
    rather than the elongated one of $\alpha$-Ti. The panels of $\Fvib$ and
    $\Fel$ tilt it here as well. The filled circle in (g),
    the minimum of the sum at 300~K, is at the strains
    $(\epsilon_a, \epsilon_c) = (+0.316, +0.229)$\%.}
\end{figure*}

In panels (e) and (f) of Fig.~\ref{fig:decomposition}, $\Fvib$ and $\Fel$
tilt the valley of $U$ when they are added to it. Panel (a) has
its minimum at the origin by construction, and panel (g) has it at the
strains $(\epsilon_a, \epsilon_c) = (+0.302, +0.202)$\%. Figure~\ref{fig:decomposition-omega} shows
the same split for $\omega$-Ti, where the curvature also comes from $U$. Its $\Fvib$ varies by $12.2$~meV
over the fitting region, and its $U$ by $13.1$~meV. The curvatures of its $U$ along
$a$ and $c$ are $8.30$ and $9.75$~eV~\AA$^{-2}$, and those of the total are
$8.01$ and $9.22$~eV~\AA$^{-2}$. The minimum of the sum is at
$(\epsilon_a, \epsilon_c) = (+0.316, +0.229)$\%. $\Fel$
differs most in size between the two phases. It varies by
$0.39$~meV over the fitting region of $\omega$-Ti and by $0.10$~meV in $\alpha$-Ti. In
both phases its variation is small compared with that of $U$, and it does
not change the curvature of the surface.

The upper rows of Figs.~\ref{fig:decomposition}
and~\ref{fig:decomposition-omega} show the same split at 0~K. At 0~K,
$\Fel$ is zero by Eq.~(\ref{eq:fel}), and $\Fvib$ is the zero-point free
energy alone. The zero-point term tilts the valley of $U$, as $\Fvib$ at
300~K does. In
$\alpha$-Ti the zero-point term varies by $4.2$~meV over the fitting region,
against $9.4$~meV at 300~K. At the minimum of the sum the size of its
gradient is $58.6$~meV~\AA$^{-1}$, and it cancels the gradient of $U$, whose
size is also $58.6$~meV~\AA$^{-1}$. The zero-point term changes the curvatures along $a$
and $c$ by only $-0.03$ and $-0.02$~eV~\AA$^{-2}$. In $\omega$-Ti the zero-point term varies by
$5.3$~meV, and the size of its gradient at the minimum is
$74.6$~meV~\AA$^{-1}$. The
tilt of the zero-point term moves the minimum of the sum away from the
origin, and this displacement is the zero-point expansion. With increasing
temperature the size of the gradient of $\Fvib$ increases, from $58.6$ to
$141$~meV~\AA$^{-1}$ in $\alpha$-Ti, so the tilt increases and the minimum
moves further.

The valleys of the two phases are similar, but the paths of their minima
are different. Both paths start away from the origin because of the
zero-point expansion, which is $(+0.098, +0.155)$\% in
$(\epsilon_a, \epsilon_c)$ for $\alpha$-Ti and $(+0.132, +0.106)$ for
$\omega$-Ti. From $T = 0$ the minimum of $\omega$-Ti moves along an almost
straight line,
$+0.184$\% in $a$ and $+0.123$ in $c$ over 0 to 300~K. The minimum of
$\alpha$-Ti moves $+0.204$\% in $a$ and only $+0.047$ in $c$, and
$c$ first decreases. For the $c$ axis of $\alpha$-Ti the zero-point
expansion, $+0.155$\%, is therefore more than three times the
expansion of that axis from 0 to 300~K, $+0.047$\%. The inset of Fig.~\ref{fig:surface}(a) shows
$c$ decreasing until 110~K and equal to its zero-temperature value again
only at 195~K. The shape of the surface at one
temperature does not show this contraction. The contraction comes from how
the surface changes with temperature, not from its curvature.

\begin{figure}
  \includegraphics{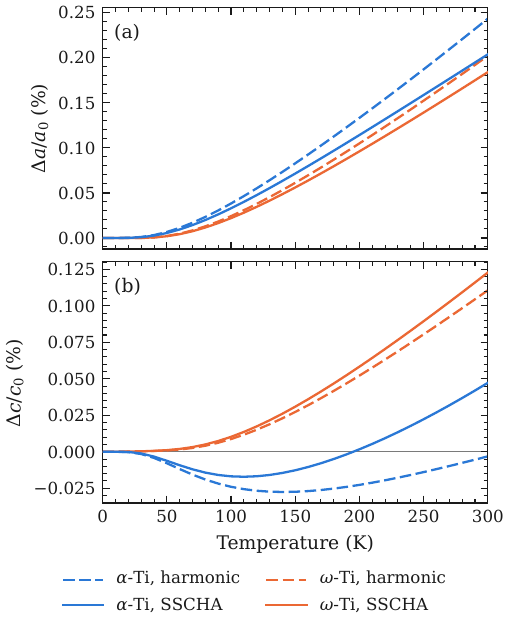}
  \caption{\label{fig:lattice}%
    Lattice parameters of $\alpha$-Ti (hcp) and $\omega$-Ti against
    temperature, each normalized to its own value at $T = 0$: (a) the $a$
    axis, (b) the $c$ axis. Dashed curves are the harmonic route and solid
    curves the SSCHA route (Sec.~\ref{sec:details}). The $c$ axis of
    $\alpha$-Ti shows negative thermal expansion at low temperature. With the SSCHA $\Fvib$ its largest contraction is
    $0.017$\% of $c$ at $110$~K, and it is equal to its
    zero-temperature value
    again at $195$~K. With the harmonic $\Fvib$ its largest contraction is
    $0.028$\% at $140$~K, and at $300$~K it is still below its
    zero-temperature value. The $c$ axis of $\omega$-Ti expands at every
    temperature.}
\end{figure}

\begin{figure}
  \includegraphics{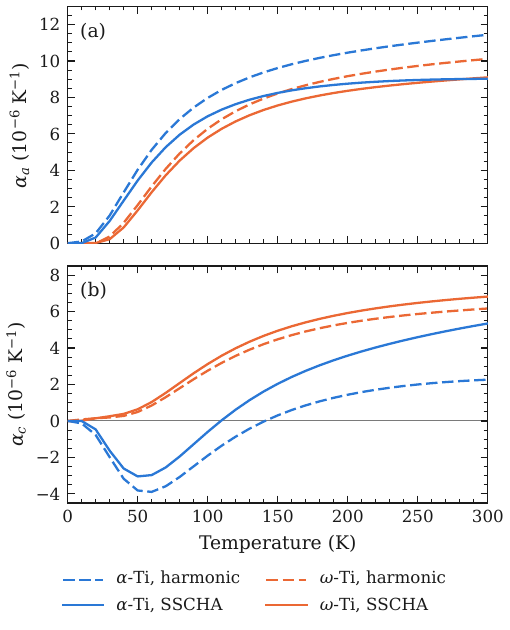}
  \caption{\label{fig:alpha}%
    Axial thermal expansion coefficients of the two phases on the harmonic
    route (dashed) and on the SSCHA route (solid), as in
    Fig.~\ref{fig:lattice}: (a) $\alpha_a$, (b) $\alpha_c$. The two phases agree in $\alpha_a$ at
    room temperature and differ qualitatively in $\alpha_c$. For $\alpha$-Ti,
    $\alpha_c$ is $-3.04 \times 10^{-6}$~K$^{-1}$ at $50$~K and crosses zero
    at $109.4$~K. For $\omega$-Ti it is positive at every temperature. Compared
    with the harmonic route, the SSCHA route gives a smaller $\alpha_a$ and a
    larger $\alpha_c$ in both phases, and the largest difference is in
    $\alpha_c$ of $\alpha$-Ti.}
\end{figure}

\subsection{Lattice parameters and axial expansions}

Figure~\ref{fig:lattice} shows the equilibrium lattice parameters of the two
phases, each normalized to its own value at $T = 0$, and
Fig.~\ref{fig:alpha} the axial thermal expansion coefficients of
Eq.~(\ref{eq:alpha}). \LatticeAtZero{} In both phases $b = a$, and hence
$\alpha_b = \alpha_a$.

The two phases differ mainly in $\alpha_c$. Their $\alpha_a$ agree to within
$0.1 \times 10^{-6}$~K$^{-1}$ at 300~K, $9.02$ for $\alpha$-Ti against
$9.10$ for $\omega$-Ti. Below about 150~K, $\alpha_a$ of $\alpha$-Ti is
larger than $\alpha_a$ of $\omega$-Ti, because it starts to increase at a
lower temperature. At 100~K it is $6.96$ against $5.80$.

The $\alpha_c$ of the two phases differ qualitatively. At low temperature
the $c$ axis of $\alpha$-Ti shows negative thermal expansion. Its
$\alpha_c$ is $-3.04 \times 10^{-6}$~K$^{-1}$ at 50~K and crosses zero at
$109.4$~K. The $\alpha_c$ of $\omega$-Ti is positive at every temperature,
and it is $3.12$ at 100~K. In the lattice parameters themselves the
largest contraction is $0.017$\% of $c$, at 110~K, and $c$ is equal to
its zero-temperature value again at 195~K. Over 0 to 300~K the axial ratio $c/a$ decreases by $0.156$\% in
$\alpha$-Ti and by $0.061$\% in $\omega$-Ti.

Compared with the harmonic route, the SSCHA route changes $\alpha_a$ and
$\alpha_c$ in opposite directions. When the
harmonic $\Fvib$ is replaced by the SSCHA one, $\alpha_a$(300~K) of
$\alpha$-Ti decreases from $11.44$ to $9.02 \times 10^{-6}$~K$^{-1}$, and
$\alpha_c$(300~K) increases from $2.26$ to $5.35$. In $\omega$-Ti,
$\alpha_a$ decreases from $10.11$ to $9.10$, and $\alpha_c$ increases from
$6.17$ to $6.82$. The two changes largely cancel in the volumetric
coefficient $2\alpha_a + \alpha_c$. For $\alpha$-Ti it changes from
$25.1$ to $23.4 \times 10^{-6}$~K$^{-1}$, $7$\%, and that of
$\omega$-Ti from $26.4$ to $25.0$, $5$\%, while $\alpha_c$ of
$\alpha$-Ti changes by a factor of $2.4$. A calculation that reports only
the volume would show a small difference between the two routes, while the
difference in the two axial coefficients is large and of opposite sign.

The SSCHA route also changes the negative thermal expansion of the $c$ axis
of $\alpha$-Ti at low temperature. With the harmonic $\Fvib$ the minimum of $\alpha_c$ is
$-3.89 \times 10^{-6}$~K$^{-1}$ and the zero crossing is at $141.0$~K. With
the SSCHA $\Fvib$ they are $-3.04$ and $109.4$~K. Both routes give negative thermal expansion.
On the harmonic route it is deeper and extends to a higher temperature.

\subsection{Cross-check against the volume path}
\label{sec:volume-path}

The diagonal of the $5 \times 5$ grid along which both $a$ and $c$ increase
is a path of nearly constant $c/a$. The ratio varies by $0.044$\%
across it for $\alpha$-Ti and $0.146$\% for $\omega$-Ti. These five
strained cells form the volume path used in the volume quasi-harmonic
approximation. Their free energies are fitted to an equation of state at
each temperature and minimized over $V$ alone, and that volume is compared
with the volume from the minimization over $a$ and $c$. Here the equation of
state is the Vinet form of phonopy's QHA module, fitted to the five strained
cells of the volume path.

The two minimizations agree. The volumetric thermal expansion coefficient at 300~K from the
equation of state is $23.42 \times 10^{-6}$~K$^{-1}$ for $\alpha$-Ti against
$23.39$ from $2\alpha_a + \alpha_c$ of the minimization over $a$ and $c$,
and $24.86$ against $25.02$
for $\omega$-Ti. Over the whole temperature range the two differ by at most
$0.035$ in $\alpha$-Ti and $0.277$ in $\omega$-Ti. The zero-temperature volumes agree to the fourth decimal in \AA$^3$ in
$\alpha$-Ti and to the second in $\omega$-Ti. The volume expansion from 0
to 300~K agrees to $0.0001$\% in $\alpha$-Ti and to
$0.0055$\% in $\omega$-Ti. The agreement is weaker in $\omega$-Ti. Its
$c/a$ varies by $0.146$\% along the volume path, against
$0.044$\% for $\alpha$-Ti, so its volume path is further from a path
of constant $c/a$.

The minimization over $a$ and $c$ contains the minimization along the
volume path as a special case. The agreement of the two minimizations is
therefore a check of the surface fit and of the minimization over $a$ and
$c$. The comparison also shows what the volume path does not give. The volume of $\alpha$-Ti
expands by $0.455$\% from 0 to 300~K in both minimizations. Over the same
temperature range the $c$ axis contracts up to 110~K and expands above it. A
calculation along the volume path holds $c/a$ nearly constant, so it cannot
show this behaviour of the $c$ axis.

\section{What each setting of the procedure changes}
\label{sec:choices}

This section measures how much each setting of the procedure changes the
axial thermal expansion coefficients. The settings measured are
\begin{itemize}
  \item the degree of the fitted surface and the number of strained cells it is fitted to;
  \item the number of terms in the Einstein fit;
  \item the method of integrating the electronic states;
  \item the random seed;
  \item the model parameters and the training set of the machine-learning
    potential.
\end{itemize}
In each measurement one setting is changed, and the other settings are kept
at the values of the reported sweep. The thermal expansion coefficient of
the $c$ axis of $\alpha$-Ti is the most sensitive quantity in this report,
and every setting is measured on it.

The random seed and the MLP itself are fixed when the SSCHA runs are made, so each of them is
measured with a separate sweep. The other settings are chosen when the sweep
is analysed. Each of them is measured by analysing the reported sweep again
with that setting changed, and any difference in the coefficients is then
caused by that setting alone.

Table~\ref{tab:choices} collects the results for every setting
except the MLP, which has its own table,
Table~\ref{tab:ladder}. The subsections explain each row and give the
quantities that are not in the tables.

\begin{table*}
  \caption{\label{tab:choices}%
    How each setting of the procedure changes the axial thermal expansion
    coefficients.
    Every entry is the difference from the coefficients of the reported
    sweep, whose settings are $l = 400$~\AA, a polynomial of total
    degree 3 over all 25 strained cells, $\Fel$ by the tetrahedron method,
    and three Einstein terms. In each row only the setting named is changed.
    Expansion coefficients are in $10^{-6}$~K$^{-1}$ at the temperature
    named. The dip is the minimum of $\alpha_c$, and $T_0$ is the
    temperature in K at which $\alpha_c$ crosses zero.
    $\omega$-Ti has no dip and no zero crossing, so the dip and $T_0$
    columns are given for $\alpha$-Ti alone.
    The seed rows come from separate sweeps, made with a different random
    seed. Every other row is a second
    analysis of the reported sweep.}
  \input{tables/choices}
\end{table*}

\subsection{The surface fit}
\label{sec:surface}

In the calculations of this report Eq.~(\ref{eq:poly}) is fitted over a
fitting region that extends less than $1$\% to either side of the
minimum in each lattice parameter. The axial thermal expansion coefficients
are logarithmic derivatives, and the Gr\"uneisen relation of
Appendix~\ref{app:derivative} is written in $\ln a$ and $\ln c$, so the
surface could be fitted in those variables instead of in $a$ and $c$. Over a
region this narrow, $\ln a$ and $1/a$ are close to straight lines in $a$, so
a polynomial in $\ln a$ and $\ln c$ spans nearly the same functions as a
polynomial in $a$ and $c$, and which of them is fitted changes the fitted
surface very little.
The
scaling of Eq.~(\ref{eq:poly}) changes only the conditioning of the fit. The
condition numbers here are computed with \texttt{numpy.linalg.cond} of
NumPy.\cite{numpy} At degree 3 the design matrix of Eq.~(\ref{eq:design})
for $\alpha$-Ti has a condition number of $2.4 \times 10^{9}$ in the lattice
parameters themselves and $8.0$ in the scaled ones. At degree 4 the two
condition numbers are $4.1 \times 10^{12}$ and $28$. The scaled values are
the same for $\omega$-Ti, because the scaled strained cells of the two
phases are the same points.

The degree of the polynomial changes the expansion coefficients of the two
phases little. The spread over degrees 2 to 4 is
$0.116 \times 10^{-6}$~K$^{-1}$ in $\alpha_c$(300~K) of $\alpha$-Ti and
$0.099$ in $\omega$-Ti.

The surface is also fitted to the free energies of fewer strained cells, on
the four $4 \times 4$ subgrids of the $5 \times 5$ grid, which are the
sixteen crosses at each corner of Fig.~\ref{fig:surface}. Each covers a
narrower fitting region than the full grid. In $\alpha$-Ti the four subgrids
change $\alpha_c$(300~K) by at most $0.07 \times 10^{-6}$~K$^{-1}$ and the
zero crossing by at most $0.91$~K, both less than the sweeps with different
MLPs in Sec.~\ref{sec:ladder} change them by. In $\omega$-Ti three of
the four change $\alpha_c$(300~K) by at most $0.08$, and the subgrid at
small $a$ and large $c$ changes it by $0.15$, which is as much as those
sweeps change it. Where the fitting region is placed therefore changes the
coefficients by at most as much as the MLP does. At degree 3 the fit determines ten
coefficients from the free energies of twenty-five strained cells, so the
sampling error of one free energy moves the fitted surface less than it
moves that free energy.

\subsection{How many Einstein terms}
\label{sec:terms}

\begin{figure*}
  \includegraphics{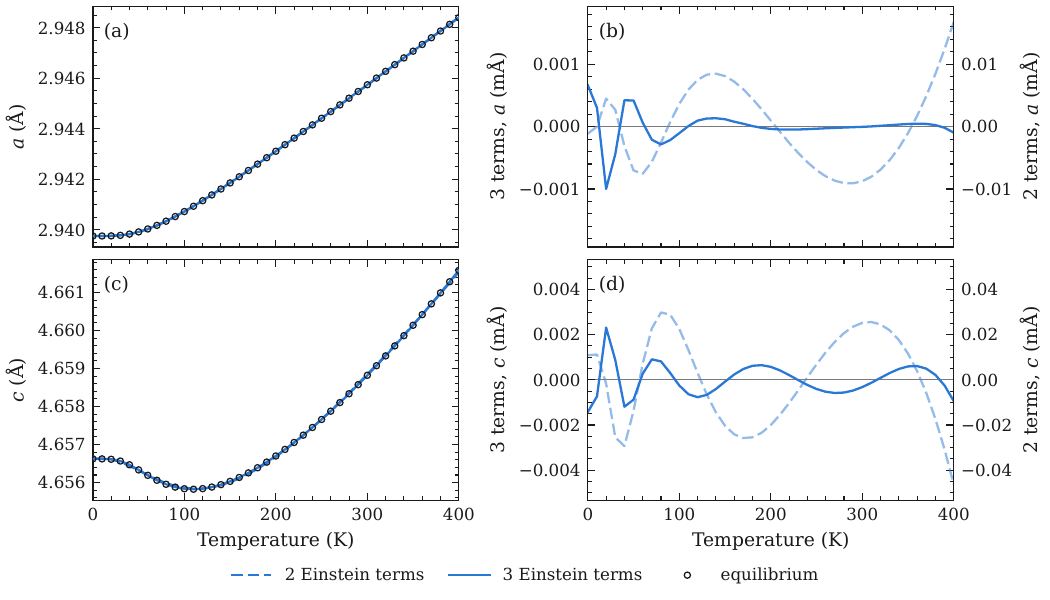}
  \caption{\label{fig:fits}%
    The equilibrium lattice parameters of $\alpha$-Ti at each temperature
    (circles) and the Einstein fits made to them, two terms dashed and three
    terms solid: (a) $a(T)$, (c) $c(T)$, with (b) and (d) the residuals of
    the fits, fit minus equilibrium value. In (b) and (d) the three-term
    residual is read on the left axis and the two-term residual on the right
    axis, whose range is ten times larger. Both fits are drawn in the colour
    used for $\alpha$-Ti in this report, and the line style distinguishes
    the two models. In (a) and (c) the equilibrium lattice parameters and
    the two fits cannot be distinguished at the scale of the lattice
    parameters. In (b) and (d) the two-term residual is a smooth
    oscillation, with a lag-one autocorrelation of $+0.90$ in $c$ and
    $+0.94$ in $a$. This oscillation is the shape of the term that the model
    is missing. The three-term residual oscillates below 80~K and is smooth
    above it.}
\end{figure*}

The number $M$ of Einstein terms in Eq.~(\ref{eq:einstein}) is chosen here
by comparing fits with two and three terms. Each fit is made to the
equilibrium lattice parameters at the 41 temperatures from 0 to 400~K, as
in Sec.~\ref{sec:details}, although the coefficients are reported only up
to 300~K. Figure~\ref{fig:fits} shows $a(T)$ and $c(T)$ of $\alpha$-Ti
with both fits drawn over them, over the whole range of the fit. At the
scale of the lattice parameters the computed values and the two fits lie on
top of each other, and only the residuals show the difference between the
two models. The residual of the two-term fit is a smooth oscillation. The
residual of the three-term fit is about thirty times smaller, and panels
(b) and (d) of Fig.~\ref{fig:fits} draw it on a scale ten times finer than
the two-term residual. The values
of a residual at neighbouring temperatures are compared by the lag-one
autocorrelation,
\begin{equation}
  R_1 = \frac{\sum_{i=1}^{N_T-1} (r_i - \bar{r}_-)(r_{i+1} - \bar{r}_+)}
             {\sqrt{\sum_{i=1}^{N_T-1} (r_i - \bar{r}_-)^2}
              \sqrt{\sum_{i=1}^{N_T-1} (r_{i+1} - \bar{r}_+)^2}},
  \label{eq:lag-one}
\end{equation}
where $r_i$ is the residual at the $i$th of the $N_T$ temperatures,
$\bar{r}_-$ is the mean of $r_1$ to $r_{N_T-1}$, and $\bar{r}_+$ is the
mean of $r_2$ to $r_{N_T}$. For the two-term fit $R_1$ is $+0.90$ in $c$ and
$+0.94$ in $a$. The two-term residual is therefore a smooth function of
temperature, and that function is the term the two-term model does not
have.

Below 80~K the three-term residual oscillates with a period of a few tens
of kelvin. Above 80~K it is smooth, with $R_1$ of $+0.85$ in $c$ and $+0.82$
in $a$. The three sweeps that differ only
in the seed (Sec.~\ref{sec:seed}) give the same three-term residual to
within $0.00001$~m\AA, so the residual does not come from the sampling. A
fit with four Einstein terms halves the residual of $c$ below 80~K and
makes the residual of $a$ six times smaller, so the oscillation below 80~K
is an error of the three-term model.

\begin{figure}
  \includegraphics{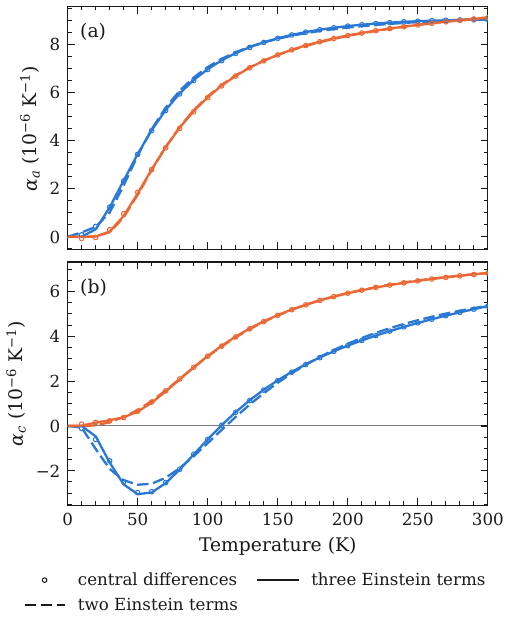}
  \caption{\label{fig:smoothing}%
    How the number of Einstein terms changes the axial thermal expansion
    coefficients:
    (a) $\alpha_a$, (b) $\alpha_c$, for $\alpha$-Ti in blue and $\omega$-Ti in
    orange. Circles are central differences of the equilibrium lattice parameters.
    Dashed curves are two Einstein terms and solid curves three. Over 0 to
    300~K the two models differ by at most $0.560 \times 10^{-6}$~K$^{-1}$ in
    $\alpha_c$ of $\alpha$-Ti, at 20~K, and by $0.212$ in its $\alpha_a$. For
    $\omega$-Ti they differ by at most $0.120$ and $0.062$.}
\end{figure}

Figure~\ref{fig:smoothing} shows the coefficients that the two models give
for both phases, together with the central differences of the equilibrium
lattice parameters. The two models differ most in $\alpha_c$ of
$\alpha$-Ti, the coefficient of the axis that contracts below 110~K and
expands above it. Over 0 to 300~K they differ there by at most
$0.560 \times 10^{-6}$~K$^{-1}$, against $0.212$ in $\alpha_a$ of
$\alpha$-Ti and $0.120$ and $0.062$ in $\omega$-Ti. In $\omega$-Ti the two models
give the quantities of Table~\ref{tab:choices} to within
$0.02 \times 10^{-6}$~K$^{-1}$. The third term is needed for the $c$ axis of
$\alpha$-Ti: with two terms the minimum of $\alpha_c$ is $0.42$ shallower
and its zero crossing is $3.56$~K higher.

The residual of a fit could be taken as an estimate of the noise in the
calculation, but neither of the two residuals is the sampling noise. Most
of the two-term residual is the error of the model
itself, and the three-term residual is the same in the three sweeps that
differ in the seed. The three-term residuals of those sweeps differ by
$0.00001$~m\AA, and this difference bounds the part of the sampling noise
that changes from one temperature to the next. Every run of a sweep draws
with the same random seed, so the sampling error at a strained cell is a
smooth function of temperature. The smooth part stays in $a(T)$ and
$c(T)$, and Section~\ref{sec:seed} measures it with sweeps that differ in
the seed.

Table~\ref{tab:einstein} gives the fitted parameters of both models for both
phases. Every three-term fit has one Einstein temperature several times
higher than 400~K, the highest temperature the Einstein fit was made to. That term
represents the nearly linear increase of the lattice
parameter at high temperature inside the fitted range.

\begin{table}
  \caption{\label{tab:einstein}%
    The Einstein fits of the reported sweep, made to the equilibrium lattice
    parameters from 0 to 400~K with two and three terms of
    Eq.~(\ref{eq:einstein}). The terms of a fit are ordered by their
    Einstein temperature, and the $m$th amplitude stands beside the $m$th
    temperature. Each term is zero at $T = 0$, so $a_0$ and $c_0$ of
    Eq.~(\ref{eq:einstein}) are the lattice parameters at $T = 0$:
    \EinsteinAtZero. The two fits ($M = 2, 3$) of a lattice parameter give
    the same $a_0$ or $c_0$ to the digits shown.}
  \input{tables/einstein}
\end{table}

\subsection{The integration of the electronic states}
\label{sec:fel-worth}

The electronic states can be integrated by summing the Fermi--Dirac
occupations over the irreducible $k$ points instead of by the tetrahedron
method. For $\alpha$-Ti this changes $\Fel$ by at most $0.019$~meV over the
25 strained cells and 41 temperatures, with a root mean square of
$0.009$~meV. For $\omega$-Ti it changes $\Fel$ by at most $0.036$~meV, with
a root mean square of $0.011$~meV. The small difference shows that the $k$-point mesh is dense
enough for the sum over the $k$ points to approximate the tetrahedron
integration.

In $\alpha$-Ti this change of at most $0.019$~meV in $\Fel$ does not change
$\alpha_c$(300~K). The row ``$k$-point sum'' of Table~\ref{tab:choices}
gives what it does change: the depth of the minimum of $\alpha_c$ moves by
$0.19 \times 10^{-6}$~K$^{-1}$ and the zero crossing by $1.9$~K. The integration method therefore changes $\alpha_c$ at low
temperature, the depth of the minimum and the zero crossing, while
$\alpha_c$(300~K) stays the same. In $\omega$-Ti the difference in $\Fel$ is
twice as large, and it changes $\alpha_c$(100~K) by
$-0.06 \times 10^{-6}$~K$^{-1}$, against $+0.13$ in $\alpha$-Ti. The two terms
of $\alpha_c$, the one driven by the entropy gradient along $a$ and the one
driven by the gradient along $c$ (Appendix~\ref{app:derivative}), cancel
less in $\omega$-Ti than in $\alpha$-Ti, so the same change in $\Fel$ moves
$\alpha_c$ less.

Leaving $\Fel$ out of the sum of Eq.~(\ref{eq:total}), the row ``left out''
of Table~\ref{tab:choices}, changes $\omega$-Ti more than $\alpha$-Ti. It
changes $\alpha_c$(300~K) of $\omega$-Ti by $-1.19 \times 10^{-6}$~K$^{-1}$,
$17$\% of it, and that of $\alpha$-Ti by $-0.15$. The integration method changes the coefficients most in
$\alpha$-Ti, where the two terms of $\alpha_c$ nearly cancel. The term $\Fel$
itself changes the coefficients most in $\omega$-Ti, where $\Fel$ is
largest.

\subsection{The random seed}
\label{sec:seed}

The random seed differs from the other settings of Table~\ref{tab:choices},
because its value has no physical meaning. The seed decides the displacements
of the supercells that the SSCHA runs draw, and free energies computed from
a finite number of drawn supercells have a sampling error. Sweeps that differ only in the seed
therefore give different axial thermal expansion coefficients. The spread of
the coefficients over those sweeps is the sampling error of the coefficients
reported here. It is obtained by repeating the sweep of each phase with
other seeds. Two of these sweeps use seeds 1001 and 1002, each shared by all
strained cells, as in the reported sweep. One more shares a seed and runs 41
iterations instead of eleven. All of them use
the MLPs, the
2500 supercells per iteration, the transient and the mesh of the reported
sweep, so the sweeps differ only in the seed and in the number of iterations.
The reported sweep and the two sweeps at seeds 1001 and 1002 are compared in
Table~\ref{tab:seeds}.

The change of the free energy with the seed is the error that
Eq.~(\ref{eq:sscha-error}) propagates. The expansion coefficients change
much less, and the reason is how the change of the free energy is
distributed over the strained cells.

Changing the seed moves the free energy of every strained cell. The move is
nearly the same at all of them, because the seed is shared, for the same
reason as the training displacements in
Sec.~\ref{sec:tdfc}. One draw of the standard normals, turned into
displacements at each strained cell with the matrices $A_{\bm{q}}$ of
Eq.~(\ref{eq:sampling-matrix}), gives displacement fields that are similar
to one another. The errors of the free energies computed from
those fields are then similar as well.

A constant added to the free energy at every strained cell shifts the
fitted surface without tilting it, so the minimum of
Eq.~(\ref{eq:minimize}) does not move. Only the part of the move that
varies from one strained cell to the next tilts the surface, and that part
is a small fraction of the move. A plane fitted to it, over the 25 strained cells at each
temperature, has a slope of $5.5$~$\mu$eV~\AA$^{-1}$ per sweep in
$\alpha$-Ti and $13.4$ in $\omega$-Ti. Divided by the
curvature of the surface, that slope moves the minimum by $0.0004$~m\AA{}
in $a$ and
$0.0002$ in $c$ for $\alpha$-Ti, and by $0.0015$ and $0.0008$~m\AA{} for
$\omega$-Ti. The $0.0002$~m\AA{} in $c$ of $\alpha$-Ti is smaller than the
residual of $c(T)$ about its three-term fit in Fig.~\ref{fig:fits}(d),
which does not change with the seed (Sec.~\ref{sec:terms}).

Table~\ref{tab:seeds} gives the reported quantities of
each sweep, with the spread over the three in its last column. No coefficient changes by more than
$0.004 \times 10^{-6}$~K$^{-1}$, and the zero crossing of $\alpha$-Ti
changes by at most $0.01$~K. The same numbers appear in
Table~\ref{tab:choices} as the rows for seeds 1001 and 1002, which have the
smallest entries in that table. These spreads also show that the spread in
the comparison of MLPs in Sec.~\ref{sec:ladder} is not sampling error. The four
training sizes of the 13,920-coefficient MLP differ by $0.165 \times 10^{-6}$~K$^{-1}$ in
$\alpha_c$(300~K). The sampling error of those sweeps is at most a twentieth of that,
so the spread comes from the MLPs themselves.

\begin{table}
  \caption{\label{tab:seeds}%
    What the random seed of the SSCHA sweep changes. Three sweeps of each
    phase share their MLPs, their 2500 supercells per iteration, their
    eleven iterations, their transient and their mesh, and differ in the seed
    alone. The reported quantities are given for each seed, with the spread
    over the three.}
  \input{tables/seeds}
\end{table}

Sharing one seed among all strained cells therefore makes the sampling error
nearly a constant over the strained cells, and a constant does not move the
minimum. A sweep should use one seed for all of its strained cells. The
three sweeps that share a seed agree to within $0.001 \times 10^{-6}$~K$^{-1}$
in $\alpha_c$(300~K).

The number of kept iterations does not change the coefficients of a sweep
that shares one seed. The sweep with 41 iterations, read with 10, 20 and 40 kept iterations, gives axial thermal
expansion coefficients that agree to within
$0.003 \times 10^{-6}$~K$^{-1}$, and four blocks of ten iterations cut out
of it scatter by $0.004$ or less. Ten kept iterations are therefore
enough.

This section measures the SSCHA sampling and not the randomness of the
training draw that the MLPs were fitted to. That draw is held fixed
throughout this report.

\subsection{The model parameters and the training set of the MLP}
\label{sec:ladder}

An SSCHA sweep evaluates the MLP for every supercell it draws, and the cost
of the sweep is mainly the cost of the MLP. This section measures how much
the model parameters and the training set change the axial thermal expansion
coefficients, by repeating the sweep with different MLPs. Each row of Table~\ref{tab:ladder} is a complete
sweep, with 25 strained cells, 41 temperatures from 0 to 400~K in steps of
10~K, 2500
supercells per iteration, eleven iterations and one random seed. Only the
MLP differs between the rows. Fifteen such sweeps were made in each
phase.

The MLPs are varied in two ways, in the model parameters and in the
training set. Which model parameters are suitable depends on the system, on
the number of elements and on the crystal structure, and no rule gives them
in advance. In practice, several sets of model parameters are tried for the
system at hand. The set chosen is one for which the thermal expansion
coefficients no longer change when the MLP is made larger, at a cost that is
acceptable, both in the time of the sweep and in the memory needed to fit
the MLPs. This section shows how that choice came out for
titanium. The upper block of Table~\ref{tab:ladder} lists the sets tried.
They range from the default model parameters of phonopy's pypolymlp
interface, which give an MLP with 781 coefficients, to a set that gives
45,680 coefficients. All of these MLPs are fitted to 180 of the 200 training
supercells, as in the reported sweep. The larger MLPs are obtained by
changing some of the default model parameters, as listed in the table. The
lower block varies the training set, with 60, 100, 140 and 180 supercells,
for the MLPs with 781 and with 13,920 coefficients.

The ridge penalty is the same at every strained cell of every sweep of
Table~\ref{tab:ladder}, and the penalty of each sweep is listed in the
table. As Sec.~\ref{sec:tdfc} explains, a penalty that changed between
strained cells would put a step into $\Fvib(\bm{x})$. For a new system, the
penalty should be chosen from fits made at all the strained cells rather
than at a single strained cell.

\begin{table*}
  \caption{\label{tab:ladder}%
    How the model parameters and the training set of the MLP change the axial
    thermal expansion coefficients. Each row is one complete sweep of both
    phases. The column of changed parameters lists the model parameters that
    differ from the default model parameters of Sec.~\ref{sec:mlp}:
    \texttt{model\_type}~4 (``type 4''), sixteen Gaussian pair parameters
    instead of eleven (``g15''), \texttt{gtinv\_order}~$n$ (``order $n$'')
    and \texttt{gtinv\_maxl} (``$l_{\max}$''). $N_\text{train}$ is the number
    of training supercells. The penalty is the ridge penalty the MLPs of that
    row were fitted at, the same in both phases. The cost is the median wall
    time of the 25 jobs of the $\alpha$-Ti sweep, one for each strained cell,
    divided by that of the sweep with the 781-coefficient MLP at 180 training
    supercells. $\omega$-Ti gives the same ratios to within $0.2$. $\alpha_a$
    and $\alpha_c$ are at 300~K in $10^{-6}$~K$^{-1}$. The upper block varies
    the model parameters and the lower block the training set.}
  \input{tables/ladder}
\end{table*}

In Table~\ref{tab:ladder}, the sweep with the 6820-coefficient MLP costs
$1.1$ times the sweep with the 781-coefficient MLP, and the sweep with the
2600-coefficient MLP costs $3.4$ times. The cost of a sweep is therefore
not proportional to the number of coefficients. The 2600-coefficient MLP is
obtained from the default model parameters by raising the number of Gaussian pair
parameters and \texttt{gtinv\_maxl}. Two pairs of rows in the table differ
in \texttt{gtinv\_maxl} alone, 1176 against 2600 coefficients and 13,920
against 27,664 coefficients. In these two pairs, the sweep with
\texttt{gtinv\_maxl} $(12, 12)$ costs $2.7$ and $2.8$ times the sweep with
$(8, 8)$. Changing \texttt{model\_type} to 4, or raising the number of
Gaussian pair parameters, multiplies the number of coefficients but
increases the cost of the sweep only by tens of per cent. Going from 781 to
13,920 coefficients makes both of these changes. It multiplies the number of
coefficients by eighteen and the cost of the sweep by $1.4$. For titanium, the cost of a sweep is
kept low by leaving \texttt{gtinv\_order} and \texttt{gtinv\_maxl} at
their default values.

The size of the training set does not change the cost of the sweep. For the
two MLPs whose training set was varied, the cost column of
Table~\ref{tab:ladder} is the same for the four training sizes, and the wall
times are within $1$\% of each other. The cost of a sweep is mainly the cost
of evaluating the MLP. The MLP is evaluated for each of the $N$ supercells
that every iteration of an SSCHA run draws, and the cost of one evaluation
does not depend on the number of training supercells the MLP was fitted to.
The number of training supercells changes a different cost, that of the
calculator. The calculator computes the energy and the forces of every
training supercell at every strained cell, before the MLPs are fitted. This
cost increases with the number of training supercells, and it is separate
from the cost of the sweep.

For the 13,920-coefficient MLP, the four training sizes of
Table~\ref{tab:ladder} give $\alpha_c$(300~K) of $\alpha$-Ti between $5.14$
and $5.30 \times 10^{-6}$~K$^{-1}$, with no trend in the number of training
supercells. The spread of these four values is $0.165 \times 10^{-6}$~K$^{-1}$. All sweeps of
Table~\ref{tab:ladder} use one random seed shared by all strained cells.
Section~\ref{sec:seed} repeats a sweep with the same MLPs at another seed,
and finds that the sampling error of $\alpha_c$(300~K) of $\alpha$-Ti is
then $0.001 \times 10^{-6}$~K$^{-1}$. The spread of $0.165 \times 10^{-6}$~K$^{-1}$ therefore comes from the MLPs, which
differ with the training set, and not from the sampling error of the SSCHA.
When two sets of model parameters in Table~\ref{tab:ladder} give values of
$\alpha_c$(300~K) that differ by less than about $0.1 \times 10^{-6}$~K$^{-1}$,
the difference may come from the training set alone, and it cannot be
attributed to the model parameters.

The seven MLPs with 2600 coefficients and more give $\alpha_c$(300~K) of
$\alpha$-Ti between $5.07$ and $5.24 \times 10^{-6}$~K$^{-1}$. The width of
this range, $0.165 \times 10^{-6}$~K$^{-1}$, is the same as the spread of
$0.165 \times 10^{-6}$~K$^{-1}$ that the training set causes for the
13,920-coefficient MLP. By the criterion above, the differences among these
seven MLPs cannot be attributed to the model parameters. The two smallest
MLPs, with 781 and 1176 coefficients, give $5.35$ and
$5.33 \times 10^{-6}$~K$^{-1}$, higher than any of the seven. The lower block
of Table~\ref{tab:ladder} shows that this difference comes from the model
parameters. There the 781- and the 13,920-coefficient MLPs are fitted to the
same four training sets. At every training size the 781-coefficient MLP
gives the higher $\alpha_c$(300~K) of $\alpha$-Ti, by $0.12$ to
$0.36 \times 10^{-6}$~K$^{-1}$. At 180 training supercells, as in the
reported sweep, the difference is $0.16 \times 10^{-6}$~K$^{-1}$, $3.1$\%.
The sweep with the 13,920-coefficient MLP costs $1.4$ times the sweep with
the 781-coefficient MLP. The sweep with the 45,680-coefficient MLP costs
$6.5$ times as much, and its $\alpha_c$(300~K) is inside the range of the
seven.

Over all fifteen sweeps of $\alpha$-Ti, the minimum of $\alpha_c$ stays
between $-3.02$ and $-3.11 \times 10^{-6}$~K$^{-1}$, and its zero crossing
stays between $108.6$ and $109.9$~K. The minimum of $\alpha_c$ and its zero crossing
therefore depend very little on the model parameters, from the
781-coefficient MLP to the 45,680-coefficient MLP. The model parameters change
$\alpha_c$(300~K) of $\alpha$-Ti by $3.1$\% and change the negative thermal
expansion at low temperature much less. For $\omega$-Ti, which has no negative thermal expansion, the
781-coefficient MLP gives an $\alpha_a$(300~K) higher than the
13,920-coefficient MLP by $0.14$ to $0.16 \times 10^{-6}$~K$^{-1}$ at every
training size, $1.7$\% at 180 training supercells.

\section{Summary}

This report has described the minimization of the free energy over the
lattice parameters as it is implemented in phonopy. The free energy is
computed at a set of strained cells. The vibrational free energy is computed
either with harmonic force constants, on the harmonic route, or with the
SSCHA and a machine-learning potential fitted at each strained cell, on the
SSCHA route. A polynomial is fitted to the free energy and minimized at each
temperature. The axial thermal expansion coefficients are obtained by
differentiating the equilibrium lattice parameters with respect to
temperature. On the SSCHA route they are differentiated through an Einstein
fit, and on the harmonic route by central differences. Both routes have been
applied to $\alpha$-Ti and $\omega$-Ti. How much each setting of the
procedure changes the coefficients has been measured by changing that one
setting and keeping the others at the values of the reported sweep.

In the measurements of Sec.~\ref{sec:choices}, the polynomial fitted to the
free energy was also fitted to the free energies of fewer strained cells, on
each of the four $4 \times 4$ subgrids of the $5 \times 5$ grid. Each
subgrid covers a narrower fitting region, and none of them is centred on
the equilibrium lattice parameters. In $\alpha$-Ti the subgrids change
$\alpha_c$(300~K) by at most $0.07 \times 10^{-6}$~K$^{-1}$, less than the
MLPs of Sec.~\ref{sec:ladder} change it. In $\omega$-Ti one subgrid changes
it by $0.15 \times 10^{-6}$~K$^{-1}$, as much as the MLPs change it.
Changing the degree of this polynomial from 2 to 4 changes
$\alpha_c$(300~K) of $\alpha$-Ti by $0.12 \times 10^{-6}$~K$^{-1}$.

In $\alpha$-Ti, whose $c$ axis contracts below 110~K, an Einstein fit
with two terms makes the minimum of $\alpha_c$
$0.42 \times 10^{-6}$~K$^{-1}$ shallower and its zero crossing $3.56$~K
higher than a fit with three terms. Three terms are therefore used.
In $\omega$-Ti, where both axes expand at every temperature, the fits with
two and with three terms give the quantities of Table~\ref{tab:choices} to
within $0.02 \times 10^{-6}$~K$^{-1}$.

Unless the $k$-point mesh is very dense, the electronic free energy has to
be integrated with the tetrahedron method, which interpolates the
eigenvalues between the $k$ points. The $k$-point mesh of this report is
dense. Even so, summing the occupations over the $k$ points instead changes
$\Fel$ of $\alpha$-Ti by at most $0.019$~meV, and this changes the depth of
the minimum of $\alpha_c$ of $\alpha$-Ti by $0.19 \times 10^{-6}$~K$^{-1}$.
The value of $\alpha_c$(300~K) stays the same.

Three sweeps that differ only in the random seed, each with one seed
shared by all strained cells, give values of $\alpha_c$(300~K) of
$\alpha$-Ti that agree to within $0.001 \times 10^{-6}$~K$^{-1}$. Their
free energies differ by several $\mu$eV, but the difference is the same
at every strained cell to within about a tenth of a $\mu$eV. A
difference that is the same at every strained cell shifts the surface
without tilting it and does not move its minimum. A sweep should
therefore use one seed for all of its strained cells.

With every MLP of Sec.~\ref{sec:ladder}, from the 781-coefficient MLP given
by the default model parameters of phonopy's pypolymlp interface to the
45,680-coefficient MLP, the minimum of $\alpha_c$ of $\alpha$-Ti is between
$-3.02$ and $-3.11 \times 10^{-6}$~K$^{-1}$, and $\alpha_c$ crosses zero
between $108.6$ and $109.9$~K. The 781-coefficient MLP gives
$\alpha_c$(300~K) of $\alpha$-Ti
$3.1$\% higher than the 13,920-coefficient MLP fitted to the same
training set, and the sweep with the 13,920-coefficient MLP is $1.4$
times as expensive. For titanium, the cost of the sweep is determined
mainly by \texttt{gtinv\_order} and \texttt{gtinv\_maxl}, not by the
number of coefficients. For another system, the model parameters have to be chosen
again, by comparing several sets of them as in Sec.~\ref{sec:ladder}.

The thermal expansion coefficient $\alpha_c$ of $\alpha$-Ti is the quantity
that the settings of Sec.~\ref{sec:choices} change most, for two reasons.
The first is the curvature of the free-energy surface. In $\alpha$-Ti the
curvature per primitive cell is $1.77$~eV~\AA$^{-2}$ along $c$ and
$12.48$~eV~\AA$^{-2}$ along $a$. An error in the free energy that differs
between strained cells tilts the fitted surface, and the minimum of the
surface is then found at different lattice parameters. Along each axis the
change is about the tilt divided by the curvature along that axis, so the
same tilt changes $c(T)$ more than $a(T)$, in angstroms. The second reason
is that $\alpha_c$ of $\alpha$-Ti is the sum of two terms of opposite sign
(Appendix~\ref{app:derivative}). At 300~K they are $-12.84$ and
$+18.19 \times 10^{-6}$~K$^{-1}$, and their sum $\alpha_c$ is
$5.35 \times 10^{-6}$~K$^{-1}$. An error of 1\% in the $+18.19$ term is therefore
an error of $3.4$\% in $\alpha_c$. An error of 1\% in the $-12.84$ term is
an error of $2.4$\% in $\alpha_c$.

At 300~K, replacing the harmonic $\Fvib$ by the SSCHA one changes the
volumetric coefficient $2\alpha_a + \alpha_c$ of $\alpha$-Ti by $7$\%, while
it changes $\alpha_c$ of $\alpha$-Ti by a factor of $2.4$. The volume
quasi-harmonic approximation along the volume path reproduces the
volumetric coefficient of $\alpha$-Ti to within $0.2$\%. The volume
therefore does not show how the two axes change, and the axial coefficients
have to be obtained from the minimization over $a$ and $c$.

\appendix

\section{The lattice derivative of the free energy}
\label{app:derivative}

The procedure of Sec.~\ref{sec:recipe} finds the equilibrium lattice
parameters $\bm{x}(T)$ at each temperature and differentiates them with
respect to temperature, through the Einstein fit on the SSCHA route and by
central differences on the harmonic route. The derivative $d\bm{x}/dT$ can
also be written with quantities at a single temperature, without comparing
two temperatures. Written in that form, it shows which quantities determine
the axial thermal expansion coefficients.

The lattice parameters $\bm{x}$ are one way of writing the strain of the unit cell. Write the unit cell as $\exp(\bm{\varepsilon})$ applied to a reference
cell, with $\varepsilon_{ij}$ the logarithmic strain tensor. The symmetry
of the crystal leaves only some components of $\bm{\varepsilon}$ free, and
$\bm{x}$ are the variables that describe those components. For a hexagonal crystal strained along
$a$ and $c$ they are
$\varepsilon_{11} = \varepsilon_{22} = \ln (a/a_0)$ and $\varepsilon_{33} =
\ln (c/c_0)$, with no shear, where $a_0$ and $c_0$ are the lattice
parameters of the reference cell. The derivation is made in $\bm{\varepsilon}$,
where it holds for any crystal, and the result is transformed to $\bm{x}$ at
the end.

Equation~(\ref{eq:minimize}) says that the equilibrium strain satisfies
\begin{equation}
  \frac{\partial F}{\partial \varepsilon_{ij}}
  \bigg|_{\bm{\varepsilon} = \bm{\varepsilon}(T)} = 0
  \label{eq:stationary}
\end{equation}
at every temperature. Differentiating Eq.~(\ref{eq:stationary}) with respect
to $T$ and using $\partial F/\partial T = -S$, where $S$ is the entropy per primitive
cell at fixed strain, gives
\begin{equation}
  \frac{\partial^2 F}{\partial \varepsilon_{ij}\, \partial \varepsilon_{kl}}
  \frac{d \varepsilon_{kl}}{dT}
  = H_{ijkl}\, \frac{d \varepsilon_{kl}}{dT}
  = \frac{\partial S}{\partial \varepsilon_{ij}} ,
  \label{eq:implicit-strain}
\end{equation}
both sides evaluated at $\bm{\varepsilon}(T)$. Here $H_{ijkl}$ is the Hessian of
$F$ with respect to the strain tensor, with one pair of indices for each of
the two strain components it is differentiated by.
Equation~(\ref{eq:implicit-strain}) is the implicit-function theorem applied
to Eq.~(\ref{eq:stationary}), and it assumes only that $F$ is stationary at
$\bm{\varepsilon}(T)$. From here on, every derivative
of $F$ or $S$ with respect to strain, and with respect to $\ln a$ and $\ln c$,
is evaluated at the equilibrium strain $\bm{\varepsilon}(T)$ of the
temperature in question, that is, at the equilibrium lattice parameters
$a(T)$ and $c(T)$.

Equation~(\ref{eq:implicit-strain}) factors the thermal expansion into two
parts, one on each side. On the left, the Hessian per unit volume is the
tensor of isothermal elastic constants, $V^{-1} H_{ijkl} = C_{ijkl}$, the
stiffness of the crystal, and the unknown is the thermal expansion tensor,
$d\varepsilon_{kl}/dT = \alpha_{kl}$. The right side contains the
temperature through the mode heat capacities, in a Gr\"uneisen sum. In a harmonic description $S$ depends on the
strain only through the phonon frequencies, and
$\partial S/\partial \omega_{\bm{q}\nu} = -C_{\bm{q}\nu} / \omega_{\bm{q}\nu}$
with $C_{\bm{q}\nu}$ the mode heat capacity, so that
\begin{equation}
  \frac{\partial S}{\partial \varepsilon_{ij}}
  = \sum_{\bm{q}\nu} C_{\bm{q}\nu}\, \gamma^{\bm{q}\nu}_{ij}
  \equiv V \lambda_{ij} ,
  \qquad
  \gamma^{\bm{q}\nu}_{ij} = - \frac{\partial \ln \omega_{\bm{q}\nu}}
  {\partial \varepsilon_{ij}} ,
  \label{eq:lambda}
\end{equation}
which defines the phonon Gr\"uneisen parameters
$\gamma^{\bm{q}\nu}_{ij}$ of Ref.~\onlinecite{wallace-thermodynamics}, a
second-rank tensor for each mode, and their heat-capacity-weighted sum
$\lambda_{ij}$. Both sides of Eq.~(\ref{eq:implicit-strain}) are then
proportional to the volume, and the volume cancels. For this reason the
volume is taken out of $\lambda_{ij}$ in Eq.~(\ref{eq:lambda}). Inverting the elastic constants gives
\begin{equation}
  \alpha_{ij} = s_{ijkl} \lambda_{kl} ,
  \label{eq:alpha-compliance}
\end{equation}
with $s_{ijkl}$ the elastic compliances, the inverse of $C_{ijkl}$.
Equation~(\ref{eq:alpha-compliance}) is the Gr\"uneisen theory of
anisotropic thermal expansion in its general
form.\cite{wallace-thermodynamics}

The third law of thermodynamics states that the entropy $S$ of the crystal
approaches a limiting value as $T \to 0$, and that this value does not
depend on the strain of the crystal. The derivative
$\partial S/\partial \varepsilon_{ij}$ therefore vanishes at $T = 0$, and so
does $\lambda_{ij}$ of Eq.~(\ref{eq:lambda}). The elastic constants
$C_{ijkl}$ are positive definite for a mechanically stable crystal, so the
compliances $s_{ijkl}$ are finite. By Eq.~(\ref{eq:alpha-compliance}) the
thermal expansion tensor $\alpha_{ij}$ is then zero at $T = 0$, and so are
$\alpha_a$ and $\alpha_c$ of Eq.~(\ref{eq:alpha}). In the harmonic
description of Eq.~(\ref{eq:lambda}) the same limit follows from the mode
heat capacities $C_{\bm{q}\nu}$, which vanish at $T = 0$.

Write the index pairs $11$, $22$ and $33$ as $a$, $b$ and $c$, so that
$\alpha_a = \alpha_{11}$, $\lambda_a = \lambda_{11}$, $s_{ab} = s_{1122}$
and so on. Then $\lambda_b = \lambda_a$ by symmetry, and
Eq.~(\ref{eq:alpha-compliance}) reads
\begin{align}
  \alpha_a &= \left( s_{aa} + s_{ab} \right) \lambda_a + s_{ac} \lambda_c ,
  \label{eq:alpha-a-compliance} \\
  \alpha_c &= 2 s_{ac} \lambda_a + s_{cc} \lambda_c ,
  \label{eq:alpha-c-compliance}
\end{align}
which are the equations usually quoted for a hexagonal crystal. The
calculation describes these two free strains through the lattice parameters
$\bm{x} = (a, c)$, with $\varepsilon_{11} = \varepsilon_{22} = \ln(a/a_0)$
and $\varepsilon_{33} = \ln(c/c_0)$. Equation~(\ref{eq:implicit-strain})
with only these two strains kept reads
\begin{equation}
  \left[ \nabla^2_{\ln\bm{x}} F \right] \frac{d\ln\bm{x}}{dT}
  = \mathsf{H}\, \frac{d\ln\bm{x}}{dT}
  = \nabla_{\ln\bm{x}} S ,
  \label{eq:implicit}
\end{equation}
where $\mathsf{H}$ is the Hessian of $F$ in $\ln a$ and $\ln c$, with
components $H_{aa}$, $H_{ac}$ and $H_{cc}$, and $d\ln\bm{x}/dT$ is
$(\alpha_a, \alpha_c)$ of Eq.~(\ref{eq:alpha}).

In this report $a$ is the one free lattice parameter for both basis
vectors $\bm{a}$ and $\bm{b}$, because $b = a$ in a hexagonal crystal
(Sec.~\ref{sec:free-parameters}). A change of $\ln a$ therefore changes two
strain components by the same amount,
$d\varepsilon_{11} = d\varepsilon_{22} = d\ln a$. With this definition a
derivative with respect to $\ln a$ is the sum of the derivatives with
respect to those two components,
\begin{align*}
  \frac{\partial S}{\partial \ln a}
  &= \frac{\partial S}{\partial \varepsilon_{11}}
  + \frac{\partial S}{\partial \varepsilon_{22}}
  = V (\lambda_a + \lambda_b) = 2 V \lambda_a , \\
  \frac{\partial S}{\partial \ln c}
  &= \frac{\partial S}{\partial \varepsilon_{33}} = V \lambda_c ,
\end{align*}
where Eq.~(\ref{eq:lambda}) and $\lambda_b = \lambda_a$ have been used. The
same holds for the Hessian: $H_{aa}$ is the sum of the four components
$H_{ijkl}$ with $ij$ and $kl$ each equal to 11 or 22, and $H_{ac}$ is the sum
of $H_{11\,33}$ and $H_{22\,33}$. Another definition, in which a change of
$\ln a$ changes the basis vector $\bm{a}$ alone, gives
$\partial S/\partial \ln a = V \lambda_a$ without the factor 2, and
$H_{aa} = H_{1111}$. Derivatives with respect to a lattice parameter taken
from other work have to be compared with the definition they were computed
with.

The surface is fitted in $a$ and $c$, so its derivatives are converted with
$d\ln a = da/a$. The entropy gradient becomes
$\partial S/\partial \ln a = a\,\partial S/\partial a$ and
$\partial S/\partial \ln c = c\,\partial S/\partial c$, and the Hessian
becomes $H_{ij} = x_i x_j\, \partial^2 F/\partial x_i \partial x_j$. In
general the Hessian in $\ln\bm{x}$ also contains a term
$\delta_{ij}\, x_i\, \partial F/\partial x_i$. That term vanishes at the
equilibrium lattice parameters, where $\partial F/\partial x_i = 0$.

The same steps apply to a crystal of another symmetry. Only the strains
that the symmetry leaves free are kept. For an orthorhombic crystal, for
example, these are $\varepsilon_{11}$, $\varepsilon_{22}$ and
$\varepsilon_{33}$, and $\bm{x} = (a, b, c)$. Equation~(\ref{eq:implicit})
has the same form in every case, with $\bm{x}$ the free lattice parameters
used in Sec.~\ref{sec:recipe}. Equations~(\ref{eq:implicit-strain}),
(\ref{eq:alpha-compliance}), (\ref{eq:alpha-c-compliance}) and
(\ref{eq:implicit}) express the same relation in different coordinates.

The axial thermal expansion coefficients of this report are not computed
from these relations. The rest of this appendix uses the relations to
interpret the coefficients.

Equation~(\ref{eq:alpha-c-compliance}) splits $\alpha_c$ into two terms, one
driven by the entropy gradient along $a$ and one driven by the entropy
gradient along $c$, and both terms can be obtained from the calculation. The
Hessian is obtained from the surface fit of Fig.~\ref{fig:surface}, and
$d\ln\bm{x}/dT$ is $(\alpha_a, \alpha_c)$ of the reported sweep. Solving
Eq.~(\ref{eq:implicit}) for $\nabla_{\ln\bm{x}} S$ therefore gives the
entropy gradient without computing an entropy, and
Eq.~(\ref{eq:alpha-c-compliance}) then gives the two terms.

\begin{table}
  \caption{\label{tab:decompose}%
    $\alpha_c$ split by Eq.~(\ref{eq:alpha-c-compliance}) into two terms, in
    $10^{-6}$~K$^{-1}$. The $a$ term is the first term of
    Eq.~(\ref{eq:alpha-c-compliance}), $2 s_{ac} \lambda_a$, and is computed
    as $(\mathsf{H}^{-1})_{ca}\, \partial S/\partial \ln a$. The $c$ term is
    the second, $s_{cc} \lambda_c$, computed as
    $(\mathsf{H}^{-1})_{cc}\, \partial S/\partial \ln c$. $\mathsf{H}$ is the
    Hessian of Eq.~(\ref{eq:implicit}), and the entropy gradients
    $\partial S/\partial \ln a$ and $\partial S/\partial \ln c$ are per
    primitive cell in $\mu$eV~K$^{-1}$. The Hessian and the gradients are
    obtained from the calculation itself, as described in the text. The
    column $\alpha_c$ is that of the reported sweep. The two terms add up to
    it by construction, because the entropy gradient is obtained from
    $(\alpha_a, \alpha_c)$ by Eq.~(\ref{eq:implicit}).}
  \input{tables/decompose}
\end{table}

Table~\ref{tab:decompose} gives this split at four temperatures. The $a$
term is negative at every temperature in both phases, for two reasons. The
off-diagonal compliance $s_{ac}$ is negative, so a stress that lengthens $a$
also shortens $c$. The derivative $\partial S/\partial \ln a$ is positive,
which means that the phonon frequencies, weighted by the mode heat
capacities, decrease on average when the crystal is strained along $a$. The
$c$ term has the sign of $\partial S/\partial \ln c$. For $\omega$-Ti,
$\partial S/\partial \ln c$ is positive at every temperature of the table,
and $\alpha_c$, the sum of the two terms, is positive at every temperature.
For $\alpha$-Ti, $\partial S/\partial \ln c$ is negative at 50~K, so both
terms of $\alpha_c$ are negative there and the negative thermal expansion has
two sources. At 100~K, $\partial S/\partial \ln c$ of $\alpha$-Ti is
positive, and the two terms, $-7.66$ and $+7.06 \times 10^{-6}$~K$^{-1}$,
nearly cancel. At the zero crossing, $109$~K, they cancel exactly.

The cancellation explains the sensitivity that Sec.~\ref{sec:choices}
measures at 300~K. At 300~K the two terms of $\alpha_c$ of $\alpha$-Ti are $-12.84$
and $+18.19$ for a sum of $5.35 \times 10^{-6}$~K$^{-1}$, so they are $2.4$ and
$3.4$ times the sum. In $\omega$-Ti the two terms are
$-5.56$ and $+12.38$ for a sum of $6.82 \times 10^{-6}$~K$^{-1}$, so they
cancel less. When two terms of opposite sign partly
cancel, a small relative error in either term gives a larger relative
error in their sum. For $\alpha_c$ of $\alpha$-Ti at 300~K the relative
error is multiplied by $2.4$ or $3.4$, depending on the term. For this reason every
setting in this report is measured on $\alpha_c$ of $\alpha$-Ti rather than
on a quantity of $\omega$-Ti.

Equation~(\ref{eq:implicit}) also shows why the $c$ axis of $\alpha$-Ti
contracts at low temperature and the $c$ axis of $\omega$-Ti does not.
Solving Eq.~(\ref{eq:implicit}) for $d\ln c/dT$, with the Hessian evaluated
at the equilibrium lattice parameters $a(T)$ and $c(T)$ of each temperature,
gives
\begin{equation*}
  \frac{d\ln c}{dT} = \frac{1}{\det \mathsf{H}}
  \left( H_{aa} \frac{\partial S}{\partial \ln c}
  - H_{ac} \frac{\partial S}{\partial \ln a} \right).
\end{equation*}
At a minimum $\det \mathsf{H} > 0$, so the sign of $d\ln c/dT$ is the sign
of the bracket. Let $\phi$ be the angle of $\nabla_{\ln\bm{x}} S$ measured
from the $\ln a$ axis. When $\partial S/\partial \ln a > 0$, as in both
phases above the lowest temperatures, the bracket changes sign where
\begin{equation}
  \tan \phi_\mathrm{c} = \frac{H_{ac}}{H_{aa}},
  \label{eq:critical-angle}
\end{equation}
so an entropy gradient pointing at an angle above $\phi_\mathrm{c}$ expands $c$
and one pointing below it contracts $c$. The angle $\phi_\mathrm{c}$ is
determined by the curvature of the free-energy surface. That curvature comes
almost entirely from $U$ (Sec.~\ref{sec:minimization}), so $\phi_\mathrm{c}$
changes little with temperature. From 10 to 300~K it changes from
$16.7^\circ$ to $17.2^\circ$ in $\alpha$-Ti and from $11.4^\circ$ to
$11.8^\circ$ in $\omega$-Ti. The direction of the gradient, in contrast, is
set by the phonons through Eq.~(\ref{eq:lambda}), and it changes with
temperature.

A second angle of the same kind gives the direction for which the expansion
is isotropic. By Eq.~(\ref{eq:implicit}), $\alpha_a = \alpha_c$ when
$d\ln\bm{x}/dT$ is along $(1, 1)$, that is, when the entropy gradient is
along $\mathsf{H}(1, 1)$. The angle of $\mathsf{H}(1, 1)$ from the $\ln a$
axis is called the isotropic angle $\phi_\mathrm{iso}$ in this report,
\begin{equation}
  \tan \phi_\mathrm{iso} = \frac{H_{ac} + H_{cc}}{H_{aa} + H_{ac}},
  \label{eq:isotropic-angle}
\end{equation}
and at 300~K it is $26.9^\circ$ for $\alpha$-Ti and $28.2^\circ$ for
$\omega$-Ti. Below $\phi_\mathrm{c}$ the $c$ axis contracts, between
$\phi_\mathrm{c}$ and $\phi_\mathrm{iso}$ it expands more slowly than $a$,
and above $\phi_\mathrm{iso}$ it expands faster.

\begin{figure}
  \includegraphics{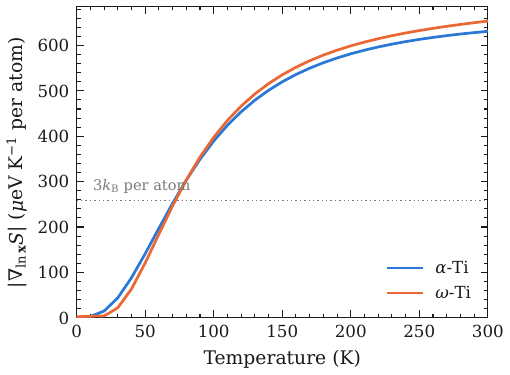}
  \caption{\label{fig:ramp}%
    The size of the entropy gradient with respect to $\ln a$ and $\ln c$,
    $|\nabla_{\ln\bm{x}} S|$, per atom, with $3\kB$ marked. Because $\ln a$
    and $\ln c$ are dimensionless, this gradient has the unit of an entropy.
    The gradient is evaluated at the equilibrium lattice parameters $a(T)$
    and $c(T)$ of each temperature. It is obtained as the temperature
    derivative of $-\nabla_{\ln\bm{x}}(\Fvib + \Fel)$ with $a$ and $c$ held
    at $a(T)$ and $c(T)$, which is the same quantity. The direction of the
    gradient is shown in Fig.~\ref{fig:ramp-circle}.}
\end{figure}

\begin{figure*}
  \includegraphics{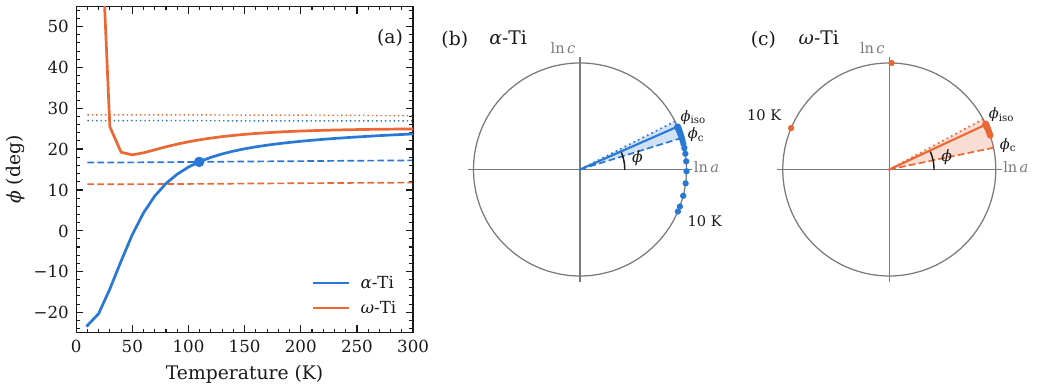}
  \caption{\label{fig:ramp-circle}%
    The direction of the entropy gradient of Fig.~\ref{fig:ramp}, in the
    strain basis. (a) The angle $\phi$ of the gradient measured from the
    $\ln a$ direction, with the critical angle $\phi_\mathrm{c}$ of
    Eq.~(\ref{eq:critical-angle}) dashed for each phase, the isotropic angle
    $\phi_\mathrm{iso}$ of Eq.~(\ref{eq:isotropic-angle}) dotted, and the
    crossing of $\alpha$-Ti marked. When the gradient of a phase is above its
    dashed line, the $c$ axis expands, and when it is below, the $c$ axis
    contracts; at the dotted line the expansion is isotropic. At the
    temperatures where a curve is above the top of (a), the gradient is
    still near zero, and its direction is the ratio of two small numbers.
    (b), (c) The same angle drawn in the $(\ln a, \ln c)$ plane, for
    $\alpha$-Ti in (b) and $\omega$-Ti in (c). Each point on the circle is
    the tip of the unit vector
    $(\partial S/\partial \ln a\,/\,|\nabla_{\ln\bm{x}} S|,\
    \partial S/\partial \ln c\,/\,|\nabla_{\ln\bm{x}} S|)$
    at one temperature, every 10~K from 10 to 300~K; 10~K is labelled for
    each phase. The solid radius is the gradient at 300~K, with its
    angle $\phi$ marked. The dashed and dotted radii are $\phi_\mathrm{c}$ and
    $\phi_\mathrm{iso}$ at 300~K, and the wedge between them is filled. A
    gradient inside the wedge expands $c$ more slowly than $a$, and a
    gradient below it contracts $c$. The gradients of both phases are just
    below $\phi_\mathrm{iso}$ at 300~K. The gradient, $\phi_\mathrm{c}$ and
    $\phi_\mathrm{iso}$ are all evaluated at $a(T)$ and $c(T)$ of each
    temperature. The numerical value of an angle depends on the basis, and
    $\phi_\mathrm{c}$ changes with it. Whether the gradient is above or
    below $\phi_\mathrm{c}$ does not depend on the basis, because it is the
    sign of $dc/dT$.}
\end{figure*}

Figure~\ref{fig:ramp} shows the size of the entropy gradient, and
Fig.~\ref{fig:ramp-circle} shows its direction. The size is the same in the two phases, to within $4$\%
above 100~K, and in both phases it has the temperature dependence of a heat
capacity. At 300~K it is $2.44$ times $3\kB$ per atom in $\alpha$-Ti and
$2.53$ times in $\omega$-Ti. By Eq.~(\ref{eq:lambda}) the gradient is a sum
of mode heat capacities weighted by Gr\"uneisen parameters, and the heat
capacity per atom approaches $3\kB$, so this ratio is of the order of the
Gr\"uneisen parameters. The direction differs between the phases.
The direction for $\omega$-Ti is above its own $\phi_\mathrm{c}$ at every
temperature, so its $c$ axis never contracts. The direction for $\alpha$-Ti
is below $\phi_\mathrm{c}$ at low temperature and crosses $\phi_\mathrm{c}$ at
$109.5$~K, close to the $109.4$~K at which $\alpha_c$ itself vanishes.

At the highest temperatures the entropy gradients of both phases point
between $\phi_\mathrm{c}$ and $\phi_\mathrm{iso}$, at a similar distance
from $\phi_\mathrm{iso}$. At 300~K the gradient of each phase is
$3.2^\circ$ below its $\phi_\mathrm{iso}$. The two phases differ in how far
$\phi_\mathrm{iso}$ is above $\phi_\mathrm{c}$ (Fig.~\ref{fig:ramp-circle}(b) and (c)). At 300~K $\phi_\mathrm{iso}$
is $9.7^\circ$ above
$\phi_\mathrm{c}$ in $\alpha$-Ti and $16.4^\circ$ above it in $\omega$-Ti.
This difference is set by the Hessian alone. By Eqs.~(\ref{eq:critical-angle}) and
(\ref{eq:isotropic-angle}), $\tan\phi_\mathrm{iso} - \tan\phi_\mathrm{c}$ is
$\det\mathsf{H} / [H_{aa}(H_{aa} + H_{ac})]$, and $\det\mathsf{H}$ is
small when the cross term $H_{ac}$ is close to the curvature along $c$,
$H_{cc}$. At 300~K, $H_{ac}$ is $0.88$ of $H_{cc}$ in $\alpha$-Ti and $0.48$
in $\omega$-Ti. Only ratios within one Hessian enter, so the comparison does
not depend on whether the energies are normalized to the primitive or the
conventional unit cell. In $\alpha$-Ti, $\phi_\mathrm{c}$ is therefore closer
to $\phi_\mathrm{iso}$, and a smaller departure of the entropy gradient from
isotropy makes the $c$ axis contract. For this reason the $c$ axis of
$\alpha$-Ti contracts and the $c$ axis of $\omega$-Ti does not.

Equation~(\ref{eq:implicit}) needs only a
second-order expansion of $F$ about the minimum. For two free
lattice parameters that expansion has six coefficients, so six strained
cells determine it. The 25 strained cells of this report over-determine it,
and the least-squares fit then averages the sampling error of $\Fvib$ over
the strained cells.

\section{Settings that follow from the shape of the unit cell}
\label{app:settings}

The strained cells differ only in their lattice parameters, and a calculator
derives several of its settings from the unit cell it is given, such as the
number of divisions of the $k$-point mesh. Such a setting is an integer, so it
does not change continuously with the lattice parameters. It can be the same
at some strained cells and differ by one at their neighbours. Where it differs, $U$ or $\Fvib$ changes by a small amount that
does not come from the change of the lattice parameters. The analysis fits a
surface to these values and finds its minimum, so this change also moves
$\bm{x}(T)$ and, through it, the axial thermal expansion coefficients.

The $k$-point mesh is one such setting. A $\Gamma$-centred mesh built from a
$k$-point spacing for the conventional unit cell is compatible with the
crystallographic point group, as Sec.~\ref{sec:recipe} requires. What
changes from one strained cell to the next is the number of divisions, which
is computed from the lengths of the reciprocal basis vectors. The divisions obtained at one strained cell
can be used at every strained cell, which keeps the sampling the same across
the grid. This matters most for the supercell calculations of the phonon
step. Their mesh is coarse, because a dense one is expensive, and on a coarse
mesh a change of one division is a large change in the sampling.

The $\bm{q}$-point mesh of the phonon calculations is given as a length $l$
in \AA, as Sec.~\ref{sec:recipe} describes. Its number of divisions is
derived from the lattice parameters in the same way, and it changes from
one strained cell to the next.

The plane-wave basis and the FFT mesh also depend on the cell. At a fixed
cutoff the number of plane waves depends on the cell, and the FFT mesh is
derived from the cell as well. Neither was fixed across the strained cells in
this report.

\section{The calculation as scripts}
\label{app:scripts}

Every step of Sec.~\ref{sec:recipe} is a phonopy command or a short script
presented in the phonopy documentation.\cite{phonopy-anisotropic-qha-doc}

\acknowledgments
\note{TODO}

\clearpage
\bibliography{refs,refs-collected}

\end{document}

%% file: tables/alpha-t0.tex
\newcommand{\LatticeAtZero}{%
With the SSCHA $\Fvib$ the lattice parameters at $T = 0$ are $a = 2.9398$, $c = 4.6566$~\AA{} for $\alpha$-Ti; $a = 4.5853$, $c = 2.8321$~\AA{} for $\omega$-Ti.}

%% file: tables/einstein-zero.tex
\newcommand{\EinsteinAtZero}{%
$a_0 = 2.9398$ and $c_0 = 4.6566$~\AA{} for $\alpha$-Ti and $a_0 = 4.5853$ and $c_0 = 2.8321$~\AA{} for $\omega$-Ti}

%% file: figures/workflow.tex
% The procedure of Sec. IV as a diagram. Hand-drawn in TikZ rather than
% generated, since it carries no data; the numbers in this manuscript all come
% from scripts/ reading data/.
%
% Left column: the chain that ends in the dataset and, with the harmonic force
% constants it stores, already gives a quasi-harmonic answer. Right column: the
% variant that replaces the vibrational free energy by an SSCHA one. Both meet
% in the analysis at the bottom.
\begin{tikzpicture}[
  font=\footnotesize,
  >={Stealth[length=1.6mm]},
  every node/.style={align=center},
  cmd/.style={draw, inner sep=3.5pt, minimum height=7mm, text width=56mm},
  calc/.style={draw, dashed, inner sep=3.5pt, minimum height=7mm,
               text width=56mm},
  data/.style={draw, rounded corners=4pt, inner sep=3.5pt, minimum height=7mm,
               text width=56mm, fill=black!4},
  stage/.style={font=\footnotesize\itshape, inner sep=1pt},
  node distance=4.2mm,
]

% ---- left column: the harmonic chain ----
\node[data] (ref) {Reference cell\\ \texttt{phonopy-init}};
\node[cmd, below=of ref] (strain)
  {$5\times5$ grid in $(a,c)$\\ \texttt{phonopy-strain-cells}};
\node[calc, below=of strain] (static)
  {Static calculation at each strained cell\\ $U$ and the electronic states};
\node[calc, below=of static] (forces)
  {Displaced supercells at each strained cell\\ forces};
\node[cmd, below=of forces] (build)
  {\texttt{phonopy-anisotropic-qha-}\\ \texttt{dataset}};
\node[data, below=of build] (dataset)
  {Dataset: strained cells, $U$, electronic\\ states, displacements, forces};

\foreach \a/\b in {ref/strain, strain/static, static/forces, forces/build,
                   build/dataset}
  \draw[->] (\a) -- (\b);

% ---- right column: the SSCHA variant ----
\node[cmd, right=30mm of strain] (draw)
  {Thermal displacements,\\ one shared draw of normals};
\node[calc, below=of draw] (tforces)
  {Calculator forces\\ training set per strained cell};
\node[cmd, below=of tforces] (mlp)
  {One potential per strained cell\\ pypolymlp};
\node[cmd, below=of mlp] (sscha)
  {SSCHA per strained cell and temperature\\ \texttt{phonopy-mlpsscha}};
\node[data, below=of sscha] (fph)
  {$\Fvib(\bm{x}; T)$ gathered\\ over the grid};

\foreach \a/\b in {draw/tforces, tforces/mlp, mlp/sscha, sscha/fph}
  \draw[->] (\a) -- (\b);

% ---- the two meet ----
\coordinate (mid) at ($(dataset.south)!0.5!(fph.south)$);
\node[cmd, below=11mm of mid, anchor=north, text width=112mm]
  (analysis)
  {\texttt{phonopy-anisotropic-qha}\\
   fit $F(a,c)$ at each $T$, minimize, fit $a(T)$ and $c(T)$};
\node[data, below=of analysis, text width=112mm] (result)
  {$a(T)$, $c(T)$, $\alpha_a(T)$, $\alpha_c(T)$};

\draw[->] (dataset.south) -- (dataset.south |- analysis.north)
  node[midway, right, stage, align=left]
  {$U$, $F_\mathrm{el}$,\\ harmonic $F_\mathrm{vib}$};
\draw[->] (fph.south) -- (fph.south |- analysis.north)
  node[midway, left, stage, align=right] {SSCHA\\ $F_\mathrm{vib}$};
\draw[->] (analysis) -- (result);

% The force constants of the dataset are what the right column starts from,
% and are also what a quasi-harmonic run would use on its own.
\draw[->] (dataset.east) -- ++(6mm,0) |- (draw.west);

\end{tikzpicture}

%% file: tables/choices.tex
% Generated by scripts/make_table_choices.py; do not edit.
\begin{ruledtabular}
\begin{tabular}{lrrrrrrrr}
 & \multicolumn{5}{c}{$\alpha$-Ti} & \multicolumn{3}{c}{$\omega$-Ti} \\
\cline{2-6}\cline{7-9}
Setting changed & $\alpha_a$ & $\alpha_c$ & $\alpha_c$ & dip & $T_0$ & $\alpha_a$ & $\alpha_c$ & $\alpha_c$ \\
 & (300 K) & (300 K) & (100 K) &  & (K) & (300 K) & (300 K) & (100 K) \\
\hline
Surface fit, degree 3 over 25 cells &  &  &  &  &  &  &  &  \\
\quad degree 2 & -0.05 & +0.07 & +0.00 & -0.06 & -0.21 & -0.07 & +0.04 & +0.05 \\
\quad degree 4 & +0.01 & -0.04 & -0.04 & -0.02 & +0.61 & +0.07 & -0.06 & -0.01 \\
\quad $4\times4$ at small $a$, small $c$ & +0.04 & -0.07 & -0.02 & +0.00 & +0.35 & +0.07 & -0.08 & -0.01 \\
\quad $4\times4$ at large $a$, small $c$ & +0.05 & -0.06 & -0.05 & -0.02 & +0.91 & +0.03 & -0.02 & -0.02 \\
\quad $4\times4$ at small $a$, large $c$ & +0.02 & +0.00 & +0.00 & -0.01 & -0.07 & +0.12 & -0.15 & -0.04 \\
\quad $4\times4$ at large $a$, large $c$ & -0.03 & +0.01 & -0.04 & -0.02 & +0.57 & +0.04 & -0.02 & -0.03 \\
Einstein fit, three terms &  &  &  &  &  &  &  &  \\
\quad two terms & +0.02 & +0.02 & -0.17 & +0.42 & +3.56 & +0.02 & +0.00 & -0.02 \\
\quad central differences & -0.00 & -0.00 & +0.00 & +0.07 & +0.05 & +0.00 & -0.01 & -0.01 \\
Electronic free energy, tetrahedron &  &  &  &  &  &  &  &  \\
\quad $k$-point sum & +0.01 & -0.01 & +0.13 & +0.19 & -1.86 & +0.04 & -0.06 & -0.06 \\
\quad left out & -0.24 & -0.15 & +0.06 & +0.02 & -0.87 & +0.81 & -1.19 & -0.70 \\
Random seed of the sweep, 1000 &  &  &  &  &  &  &  &  \\
\quad seed 1001 & -0.00 & -0.00 & -0.00 & -0.00 & +0.01 & -0.00 & +0.00 & +0.00 \\
\quad seed 1002 & -0.00 & -0.00 & -0.00 & -0.00 & +0.00 & +0.00 & -0.00 & -0.00 \\
\end{tabular}
\end{ruledtabular}

%% file: tables/einstein.tex
% Generated by scripts/make_table_einstein.py; do not edit.
\begin{ruledtabular}
\begin{tabular}{llll}
Phase & $M$ & $\theta_m$ (K) & $A_m$ ($10^{-5}$~\AA~K$^{-1}$) \\
\hline
$\alpha$-Ti, $a$ & 2 & 34, 192 & $+0.13$, $+2.62$ \\
 & 3 & 125, 246, 1413 & $+1.18$, $+1.64$, $-0.26$ \\
$\alpha$-Ti, $c$ & 2 & 80, 449 & $-1.57$, $+4.87$ \\
 & 3 & 123, 356, 1399 & $-2.65$, $+5.40$, $+1.42$ \\
$\omega$-Ti, $a$ & 2 & 243, 1200 & $+4.28$, $+0.45$ \\
 & 3 & 233, 426, 2150 & $+3.88$, $+0.52$, $+1.12$ \\
$\omega$-Ti, $c$ & 2 & 162, 386 & $+0.34$, $+1.83$ \\
 & 3 & 74, 354, 1317 & $+0.11$, $+2.01$, $+0.12$ \\
\end{tabular}
\end{ruledtabular}

%% file: tables/seeds.tex
% Generated by scripts/make_table_seeds.py; do not edit.
\begin{ruledtabular}
\begin{tabular}{llrrrr}
 & & 1000 & 1001 & 1002 & spread \\
\hline
\multicolumn{6}{l}{%
  the reported quantities, $10^{-6}$~K$^{-1}$ and K} \\
$\alpha$-Ti & $\alpha_a$(100~K) & 6.963 & 6.963 & 6.963 & 0.001 \\
 & $\alpha_c$(100~K) & -0.602 & -0.602 & -0.602 & 0.001 \\
 & $\alpha_a$(300~K) & 9.022 & 9.021 & 9.020 & 0.002 \\
 & $\alpha_c$(300~K) & 5.348 & 5.347 & 5.348 & 0.001 \\
 & dip of $\alpha_c$ & -3.044 & -3.045 & -3.045 & 0.001 \\
 & $T_0$ (K) & 109.44 & 109.45 & 109.44 & 0.01 \\
$\omega$-Ti & $\alpha_a$(100~K) & 5.795 & 5.792 & 5.795 & 0.003 \\
 & $\alpha_c$(100~K) & 3.121 & 3.122 & 3.120 & 0.002 \\
 & $\alpha_a$(300~K) & 9.099 & 9.094 & 9.099 & 0.004 \\
 & $\alpha_c$(300~K) & 6.823 & 6.825 & 6.821 & 0.004 \\
\end{tabular}
\end{ruledtabular}

%% file: tables/ladder.tex
% Generated by scripts/make_table_ladder.py; do not edit.
\begin{ruledtabular}
\begin{tabular}{rlrrrrrrr}
 & & & & & \multicolumn{2}{c}{$\alpha$-Ti} & \multicolumn{2}{c}{$\omega$-Ti} \\
\cline{6-7}\cline{8-9}
Coefficients & Changed parameters & $N_\text{train}$ & Penalty & Cost & $\alpha_a$ & $\alpha_c$ & $\alpha_a$ & $\alpha_c$ \\
\hline
781 & -- & 180 & $10^{-3}$ & 1.0 & 9.02 & 5.35 & 9.10 & 6.82 \\
1,176 & g15 & 180 & $10^{-3}$ & 1.3 & 9.00 & 5.33 & 9.07 & 6.80 \\
2,600 & g15, $l_{\max}(12,12)$ & 180 & $10^{-3}$ & 3.4 & 9.09 & 5.18 & 9.04 & 6.80 \\
3,848 & g15, order 4, $l_{\max}(16,12,4)$ & 180 & $10^{-3}$ & 4.8 & 9.06 & 5.24 & 9.03 & 6.81 \\
6,820 & type 4 & 180 & $10^{-3}$ & 1.1 & 9.13 & 5.07 & 8.98 & 6.84 \\
13,920 & type 4, g15 & 180 & $10^{-3}$ & 1.4 & 9.09 & 5.19 & 8.94 & 6.88 \\
22,495 & type 4, order 6, $l_{\max}(16,12,4,1,1)$ & 180 & $10^{0}$ & 3.6 & 9.09 & 5.16 & 9.01 & 6.80 \\
27,664 & type 4, g15, $l_{\max}(12,12)$ & 180 & $10^{0}$ & 4.0 & 9.09 & 5.18 & 9.00 & 6.81 \\
45,680 & type 4, g15, order 6, $l_{\max}(16,12,4,1,1)$ & 180 & $10^{0}$ & 6.5 & 9.10 & 5.16 & 8.99 & 6.82 \\
\hline
781 & -- & 60 & $10^{-3}$ & 1.0 & 8.92 & 5.57 & 9.09 & 6.81 \\
781 & -- & 100 & $10^{-3}$ & 1.0 & 8.99 & 5.43 & 9.11 & 6.78 \\
781 & -- & 140 & $10^{-3}$ & 1.0 & 9.02 & 5.38 & 9.10 & 6.81 \\
13,920 & type 4, g15 & 60 & $10^{-3}$ & 1.4 & 9.01 & 5.21 & 8.94 & 6.93 \\
13,920 & type 4, g15 & 100 & $10^{-3}$ & 1.4 & 9.01 & 5.30 & 8.97 & 6.83 \\
13,920 & type 4, g15 & 140 & $10^{-3}$ & 1.4 & 9.12 & 5.14 & 8.95 & 6.90 \\
\end{tabular}
\end{ruledtabular}

%% file: tables/decompose.tex
% Generated by scripts/make_table_decompose.py; do not edit.
\begin{ruledtabular}
  \begin{tabular}{crrrrr}
    $T$ (K) & $\partial S/\partial \ln a$ & $\partial S/\partial \ln c$
    & $a$ term & $c$ term & $\alpha_c$ \\
    \hline
    \multicolumn{6}{c}{$\alpha$-Ti} \\
    50 & 281 & $-7$ & $-2.81$ & $-0.24$ & $-3.04$ \\
    100 & 752 & 210 & $-7.66$ & $7.06$ & $-0.60$ \\
    200 & 1079 & 434 & $-11.50$ & $15.09$ & $3.58$ \\
    300 & 1156 & 508 & $-12.84$ & $18.19$ & $5.35$ \\
    \multicolumn{6}{c}{$\omega$-Ti} \\
    50 & 334 & 112 & $-0.96$ & $1.59$ & $0.64$ \\
    100 & 1111 & 444 & $-3.21$ & $6.33$ & $3.12$ \\
    200 & 1637 & 743 & $-4.92$ & $10.84$ & $5.92$ \\
    300 & 1779 & 827 & $-5.56$ & $12.38$ & $6.82$ \\
  \end{tabular}
\end{ruledtabular}

%% file: refs-collected.bib
@article{phonopy,
  journal = {Scr. Mater.},
  year    = {2015},
  title   = {First principles phonon calculations in materials science},
  author  = {Togo, A. and Tanaka, I.},
  pages   = {1--5},
  volume  = {108},
  month   = {Nov}
}

@article{phonopy-phono3py,
  author  = {Togo ,Atsushi},
  title   = {First-principles Phonon Calculations with Phonopy and Phono3py},
  journal = {J. Phys. Soc. Jpn.},
  volume  = {92},
  number  = {1},
  pages   = {012001},
  year    = {2023},
  doi     = {10.7566/JPSJ.92.012001}
}

@article{phonopy-QHA-max,
  title     = {First-principles phonon calculations of thermal expansion in {Ti$_{3}$SiC$_{2}$}, {Ti$_{3}$AlC$_{2}$}, and {Ti$_{3}$GeC$_{2}$}},
  author    = {Togo, Atsushi and Chaput, Laurent and Tanaka, Isao and Hug, Gilles},
  journal   = {Phys. Rev. B},
  volume    = {81},
  issue     = {17},
  pages     = {174301},
  numpages  = {6},
  year      = {2010},
  month     = {May},
  publisher = {American Physical Society},
  doi       = {10.1103/PhysRevB.81.174301}
}

@article{VASP-Kresse-1995,
  journal = {J. Non-Cryst. Solids},
  year    = {1995},
  title   = {Ab-Initio Molecular-Dynamics For Liquid-Metals},
  author  = {Kresse, G},
  pages   = {222--229},
  volume  = {193},
  month   = {Dec}
}

@article{VASP-Kresse-1996,
  journal = {Comput. Mater. Sci.},
  year    = {1996},
  title   = {Efficiency of ab-initio total energy calculations for metals and semiconductors using a plane-wave basis set},
  author  = {Kresse, G and Furthm\"{u}ller, J},
  pages   = {15--50},
  volume  = {6},
  issue   = {1},
  month   = {Jul}
}

@article{VASP-Kresse-1999,
  journal = {Phys. Rev. B},
  year    = {1999},
  title   = {From ultrasoft pseudopotentials to the projector augmented-wave method},
  author  = {Kresse, G and Joubert, D},
  pages   = {1758--1775},
  volume  = {59},
  issue   = {3},
  month   = {Jan}
}

@article{PAW-Blochl-1994,
  journal = {Phys. Rev. B},
  year    = {1994},
  title   = {Projector Augmented-Wave Method},
  author  = {Bl\"{o}chl, P. E.},
  pages   = {17953--17979},
  volume  = {50},
  issue   = {24},
  month   = {Dec}
}

@article {Perdew-PBE-1996,
     Journal = {Phys. Rev. Lett.},
     Year = {1996},
     Title = {Generalized gradient approximation made simple},
     Author = {Perdew, J. P. and Burke, K and Ernzerhof, M},
     Pages = {3865--3868},
     Volume = {77},
     Issue = {18},
     Month = {Oct}
}

@article{Blochl-tetrahedron-1994,
  title     = {Improved tetrahedron method for {Brillouin}-zone integrations},
  author    = {Bl\"ochl, Peter E. and Jepsen, O. and Andersen, O. K.},
  journal   = {Phys. Rev. B},
  volume    = {49},
  issue     = {23},
  pages     = {16223--16233},
  numpages  = {0},
  year      = {1994},
  month     = {Jun},
  publisher = {American Physical Society},
  doi       = {10.1103/PhysRevB.49.16223}
}

@article{Errea-SSCHA-2013,
  title     = {First-Principles Theory of Anharmonicity and the Inverse Isotope Effect in Superconducting Palladium-Hydride Compounds},
  author    = {Errea, Ion and Calandra, Matteo and Mauri, Francesco},
  journal   = {Phys. Rev. Lett.},
  volume    = {111},
  issue     = {17},
  pages     = {177002},
  numpages  = {5},
  year      = {2013},
  month     = {Oct},
  publisher = {American Physical Society},
  doi       = {10.1103/PhysRevLett.111.177002}
}

@article{Errea-SSCHA-2014,
  title     = {Anharmonic free energies and phonon dispersions from the stochastic self-consistent harmonic approximation: Application to platinum and palladium hydrides},
  author    = {Errea, Ion and Calandra, Matteo and Mauri, Francesco},
  journal   = {Phys. Rev. B},
  volume    = {89},
  issue     = {6},
  pages     = {064302},
  numpages  = {16},
  year      = {2014},
  month     = {Feb},
  publisher = {American Physical Society},
  doi       = {10.1103/PhysRevB.89.064302}
}

@article{Bianco-SSCHA-2017,
  title     = {Second-order structural phase transitions, free energy curvature, and temperature-dependent anharmonic phonons in the self-consistent harmonic approximation: Theory and stochastic implementation},
  author    = {Bianco, Raffaello and Errea, Ion and Paulatto, Lorenzo and Calandra, Matteo and Mauri, Francesco},
  journal   = {Phys. Rev. B},
  volume    = {96},
  issue     = {1},
  pages     = {014111},
  numpages  = {26},
  year      = {2017},
  month     = {Jul},
  publisher = {American Physical Society},
  doi       = {10.1103/PhysRevB.96.014111}
}

@article{Monacelli-SSCHA-2021,
  doi       = {10.1088/1361-648x/ac066b},
  year      = {2021},
  month     = {jul},
  publisher = {{IOP} Publishing},
  volume    = {33},
  number    = {36},
  pages     = {363001},
  author    = {Monacelli, Lorenzo and Bianco, Raffaello and Cherubini, Marco and Calandra, Matteo and Errea, Ion and Mauri, Francesco},
  title     = {The stochastic self-consistent harmonic approximation: calculating vibrational properties of materials with full quantum and anharmonic effects},
  journal   = {J. Phys. Condensed Matter}
}

@article{Tadano-2015,
  title     = {Self-consistent phonon calculations of lattice dynamical properties in cubic {SrTiO$_{3}$} with first-principles anharmonic force constants},
  author    = {Tadano, T. and Tsuneyuki, S.},
  journal   = {Phys. Rev. B},
  volume    = {92},
  issue     = {5},
  pages     = {054301},
  numpages  = {10},
  year      = {2015},
  month     = {Aug},
  publisher = {American Physical Society},
  doi       = {10.1103/PhysRevB.92.054301}
}

@article{Togo-IXS-KCl-2022,
  doi       = {10.1088/1361-648x/ac7b01},
  year      = 2022,
  month     = {jul},
  publisher = {{IOP} Publishing},
  volume    = {34},
  number    = {36},
  pages     = {365401},
  author    = {Togo, Atsushi  and Hayashi, Hiroyuki  and Tadano, Terumasa and Tsutsui, Satoshi and Tanaka, Isao},
  title     = {{LO}-mode phonon of {KCl} and {NaCl} at 300{\hspace{0.167em}}{K} by inelastic x-ray scattering measurements and first principles calculations},
  journal   = {J. Phys. Condens. Matter}
}


%% file: refs.bib
@article{pypolymlp,
  author  = {Seko, Atsuto},
  title   = {Tutorial: Systematic development of polynomial machine learning
             potentials for elemental and alloy systems},
  journal = {J. Appl. Phys.},
  volume  = {133},
  number  = {1},
  pages   = {011101},
  year    = {2023},
  doi     = {10.1063/5.0129045}
}

@article{symfc,
  title   = {Projector-based efficient estimation of force constants},
  author  = {Seko, Atsuto and Togo, Atsushi},
  journal = {Phys. Rev. B},
  volume  = {110},
  issue   = {21},
  pages   = {214302},
  year    = {2024},
  doi     = {10.1103/PhysRevB.110.214302}
}

@article{phonopy-implementation,
  author  = {Togo, Atsushi and Chaput, Laurent and Tadano, Terumasa and
             Tanaka, Isao},
  title   = {Implementation strategies in phonopy and phono3py},
  journal = {J. Phys.: Condens. Matter},
  volume  = {35},
  number  = {35},
  pages   = {353001},
  year    = {2023},
  doi     = {10.1088/1361-648X/acd831}
}

@article{chatterji-einstein-gruneisen,
  author  = {Chatterji, Tapan and Hansen, Thomas C.},
  title   = {Magnetoelastic effects in {J}ahn--{T}eller distorted
             {CrF$_2$} and {CuF$_2$} studied by neutron powder diffraction},
  journal = {J. Phys.: Condens. Matter},
  volume  = {23},
  number  = {27},
  pages   = {276007},
  year    = {2011},
  doi     = {10.1088/0953-8984/23/27/276007}
}

@book{wallace-thermodynamics,
  author    = {Wallace, Duane C.},
  title     = {Thermodynamics of Crystals},
  publisher = {John Wiley \& Sons},
  address   = {New York},
  year      = {1972}
}

@article{angel-eosfit,
  author  = {Angel, Ross J. and Gonzalez-Platas, Javier and Alvaro, Matteo},
  title   = {{EosFit7c} and a {F}ortran module (library) for equation of
             state calculations},
  journal = {Z. Kristallogr.},
  volume  = {229},
  number  = {5},
  pages   = {405--419},
  year    = {2014}
}

@article{allan-zsisa,
  author  = {Allan, N. L. and Barron, T. H. K. and Bruno, J. A. O.},
  title   = {The zero static internal stress approximation in lattice
             dynamics, and the calculation of isotope effects on molar
             volumes},
  journal = {J. Chem. Phys.},
  volume  = {105},
  number  = {18},
  pages   = {8300--8303},
  year    = {1996},
  doi     = {10.1063/1.472684}
}

@article{gong-beryllium,
  author  = {Gong, Xuejun and Dal Corso, Andrea},
  title   = {High temperature and pressure thermoelasticity of hcp metals
             from ab initio quasi-harmonic free energy calculations: the
             beryllium case},
  journal = {Phys. Rev. B},
  volume  = {110},
  pages   = {094109},
  year    = {2024}
}

@article{gong-osmium,
  author  = {Gong, Xuejun and Dal Corso, Andrea},
  title   = {High pressure and temperature thermoelasticity of hcp osmium
             from ab initio quasi-harmonic theory},
  journal = {Phys. Rev. B},
  volume  = {112},
  pages   = {024103},
  year    = {2025}
}

@article{rostami-anisotropic,
  author  = {Rostami, Samare and Giantomassi, Matteo and Gonze, Xavier},
  title   = {Anisotropic temperature-dependent lattice parameters and elastic
             constants from first principles},
  journal = {npj Comput. Mater.},
  volume  = {11},
  pages   = {271},
  year    = {2025},
  doi     = {10.1038/s41524-025-01765-5}
}

@misc{phonopy-anisotropic-qha-doc,
  title = {Anisotropic thermal expansion from free-energy minimization},
  howpublished = {Phonopy documentation},
  note = {\url{https://phonopy.github.io/phonopy/aniso-thermal-expansion.html}},
  year = {2026},
}

@article{van-Roekeghem-2020,
  title    = {Quantum Self-Consistent Ab-Initio Lattice Dynamics},
  journal  = {Comput. Phys. Commun.},
  volume   = {263},
  pages    = {107945},
  year     = {2021},
  issn     = {0010-4655},
  doi      = {https://doi.org/10.1016/j.cpc.2021.107945},
  author   = {{van Roekeghem}, Ambroise and Carrete, Jes\'us and Mingo, Natalio}
}

@article{scipy,
  author  = {Virtanen, Pauli and Gommers, Ralf and Oliphant, Travis E. and
             Haberland, Matt and Reddy, Tyler and Cournapeau, David and
             Burovski, Evgeni and Peterson, Pearu and Weckesser, Warren and
             Bright, Jonathan and others},
  title   = {{SciPy} 1.0: fundamental algorithms for scientific computing in
             {Python}},
  journal = {Nat. Methods},
  volume  = {17},
  pages   = {261--272},
  year    = {2020},
  doi     = {10.1038/s41592-019-0686-2}
}

@article{numpy,
  author  = {Harris, Charles R. and Millman, K. Jarrod and
             van der Walt, St\'efan J. and Gommers, Ralf and
             Virtanen, Pauli and Cournapeau, David and Wieser, Eric and
             Taylor, Julian and Berg, Sebastian and Smith, Nathaniel J. and
             others},
  title   = {Array programming with {NumPy}},
  journal = {Nature},
  volume  = {585},
  pages   = {357--362},
  year    = {2020},
  doi     = {10.1038/s41586-020-2649-2}
}

@book{nocedal-wright,
  author    = {Nocedal, Jorge and Wright, Stephen J.},
  title     = {Numerical Optimization},
  edition   = {2nd},
  publisher = {Springer},
  address   = {New York},
  year      = {2006},
  doi       = {10.1007/978-0-387-40065-5}
}

@article{levenberg,
  author  = {Levenberg, Kenneth},
  title   = {A method for the solution of certain non-linear problems in
             least squares},
  journal = {Q. Appl. Math.},
  volume  = {2},
  pages   = {164--168},
  year    = {1944},
  doi     = {10.1090/qam/10666}
}

@article{marquardt,
  author  = {Marquardt, Donald W.},
  title   = {An algorithm for least-squares estimation of nonlinear
             parameters},
  journal = {J. Soc. Ind. Appl. Math.},
  volume  = {11},
  pages   = {431--441},
  year    = {1963},
  doi     = {10.1137/0111030}
}

@article{PhysRevB.105.064112,
  title = {Anharmonic {Gr\"uneisen} theory based on self-consistent phonon theory: Impact of phonon-phonon interactions neglected in the quasiharmonic theory},
  author = {Masuki, Ryota and Nomoto, Takuya and Arita, Ryotaro and Tadano, Terumasa},
  journal = {Phys. Rev. B},
  volume = {105},
  issue = {6},
  pages = {064112},
  year = {2022},
  month = {Feb},
  publisher = {American Physical Society},
  doi = {10.1103/PhysRevB.105.064112}
}

@article{PhysRevB.107.134119,
  title = {Full optimization of quasiharmonic free energy with an anharmonic lattice model: Application to thermal expansion and pyroelectricity of wurtzite {GaN} and {ZnO}},
  author = {Masuki, Ryota and Nomoto, Takuya and Arita, Ryotaro and Tadano, Terumasa},
  journal = {Phys. Rev. B},
  volume = {107},
  issue = {13},
  pages = {134119},
  year = {2023},
  month = {Apr},
  publisher = {American Physical Society},
  doi = {10.1103/PhysRevB.107.134119}
}

@article{PhysRevB.106.224104,
  title = {Ab initio structural optimization at finite temperatures based on anharmonic phonon theory: Application to the structural phase transitions of {BaTiO$_{3}$}},
  author = {Masuki, Ryota and Nomoto, Takuya and Arita, Ryotaro and Tadano, Terumasa},
  journal = {Phys. Rev. B},
  volume = {106},
  issue = {22},
  pages = {224104},
  numpages = {26},
  year = {2022},
  month = {Dec},
  publisher = {American Physical Society},
  doi = {10.1103/PhysRevB.106.224104},
  url = {https://link.aps.org/doi/10.1103/PhysRevB.106.224104}
}

@article{MacDonald-tetrahedron-1979,
  author  = {MacDonald, A. H. and Vosko, S. H. and Coleridge, P. T.},
  title   = {Extensions of the tetrahedron method for evaluating spectral properties of solids},
  journal = {J. Phys. C: Solid State Phys.},
  volume  = {12},
  number  = {15},
  pages   = {2991},
  year    = {1979},
  doi     = {10.1088/0022-3719/12/15/008}
}
